\documentclass[a4paper,12pt]{article}
\pdfoutput=1
\usepackage{amssymb,amsmath,bm}
\usepackage{color}
\usepackage{slashed}
\usepackage{cite}
\usepackage[latin1]{inputenc}
\usepackage{amsfonts}
\usepackage{lscape}
\usepackage{amsthm}
\usepackage{booktabs}
\usepackage{array}
\usepackage{rotating}
\usepackage{float}
\usepackage{multirow}
\usepackage{graphicx,rotating,amsmath,multirow,xcolor,colortbl,xcolor}
\usepackage{hyperref,pdflscape,slashed}
\definecolor{rosso}{cmyk}{0,1,1,0.4}
\definecolor{rossos}{cmyk}{0,1,1,0.55}
\definecolor{rossoc}{cmyk}{0,1,1,0.2}
\definecolor{blu}{cmyk}{1,1,0,0.3}
\definecolor{blus}{cmyk}{1,1,0,0.6}
\definecolor{bluc}{cmyk}{1,1,0,0.1}
\definecolor{verde}{cmyk}{0.92,0,0.59,0.25}
\definecolor{verdec}{cmyk}{0.92,0,0.59,0.15}
\definecolor{verdes}{cmyk}{0.92,0,0.59,0.4}
\definecolor{Gray}{gray}{0.95}

\font\tenrsfs=rsfs10 at 12pt
\font\sevenrsfs=rsfs7
\font\fiversfs=rsfs5
\newfam\rsfsfam
\textfont\rsfsfam=\tenrsfs
\scriptfont\rsfsfam=\sevenrsfs
\scriptscriptfont\rsfsfam=\fiversfs
\def\mathscr#1{{\fam\rsfsfam\relax#1}}

\usepackage[small,bf]{caption}
\hypersetup{colorlinks,bookmarksopen,bookmarksnumbered,
linkcolor=blus,pdfstartview=FitH,urlcolor=rossos,citecolor=verde}

\newcommand{\lsim}{\stackrel{<}{_\sim}}

\newcommand{\hc}{{\rm h.c.}}

\newcommand{\be}{\begin{equation}}
\newcommand{\ee}{\end{equation}}
\newcommand{\bea}{\begin{eqnarray}}
\newcommand{\eea}{\end{eqnarray}}
\newcommand{\beq}{\begin{equation}}
\newcommand{\eeq}{\end{equation}}
\newcommand{\beqa}{\begin{eqnarray}}
\newcommand{\eeqa}{\end{eqnarray}}

\def\eq#1{eq.~(\ref{#1})}
\def\fig#1{fig.~\ref{#1}}
\def\sect#1{sect.~\ref{#1}}

\def\Eq#1{eq.~(\ref{#1})}
\def\Fig#1{fig.~\ref{fig:#1}}
\def\Sec#1{sect.~\ref{#1}}
\newcommand{\Appf}[1]{appendix~\ref{sec:#1}}

\newcommand{\sectioneq}[1]{\section{#1}\setcounter{equation}{0}}

\begin{document}

{\hfill CERN-TH-2016-223}

\vspace{2cm}

\begin{center}
\boldmath

{\textbf{\LARGE A Clockwork Theory
}}
\unboldmath

\bigskip

\vspace{0.4 truecm}

{\bf Gian F. Giudice} and {\bf Matthew McCullough}
 \\[5mm]

{\it CERN, Theoretical Physics Department, Geneva, Switzerland}\\[2mm]

\vspace{2cm}

{\bf Abstract }
\end{center}

\begin{quote}
The clockwork is a mechanism for generating light particles with exponentially suppressed interactions in theories which contain no small parameters at the fundamental level. We develop a general description of the clockwork mechanism valid for scalars, fermions, gauge bosons, and gravitons. This mechanism can be implemented with a discrete set of new fields or, in its continuum version, through an extra spatial dimension. In both cases the clockwork emerges as a useful tool for model-building applications.  Notably, the continuum clockwork offers a solution to the Higgs naturalness problem, which turns out to be the same as in linear dilaton duals of Little String Theory.  We also elucidate the similarities and differences of the continuum clockwork with large extra dimensions and warped spaces. All clockwork models, in the discrete and continuum, exhibit novel phenomenology with a distinctive spectrum of closely spaced resonances.
\end{quote}

\thispagestyle{empty}
\vfill

\newpage

\tableofcontents
%\newpage
%%%%%%%%%%%%%%%%%%%%%%%%%%%%%%%%%%%%%%%%%

\sectioneq{Introduction}
\label{sec:intro}
In theories with only local interactions, new physics effects can be described at low energy as non-renormalisable operators involving Standard Model (SM) fields. These operators necessarily involve interaction scales, which are often identified as the energy at which new dynamics must take place. Examples are: the Weinberg operator, whose scale is usually associated with the right-handed neutrino mass; the axion decay constant, associated with the Peccei-Quinn (PQ) breaking dynamics; the scale of baryon-number violating operators, associated with the mass of GUT particles; the Planck scale, associated with the energy at which quantum gravity emerges. This association between interaction scales and the mass scale of the UV-completion is fallacious or, at least, is based on a hidden assumption.

As discussed in detail in \sect{sec:massesscalescouplings}, scales and masses are intrinsically different physical quantities, carrying different units of measure. Commensurable quantities are masses and the product of scales and couplings. Therefore, if couplings are ${\mathcal O}(1)$, when measured in natural units, then the distinction between masses and scales has little consequence. In this case, the association between the interaction scale and the energy at which new particles must enter, although not formally correct, works in practice. The situation is very different in presence of couplings which are small, in natural units, as the dynamics associated with an interaction scale could occur at much smaller energies.

These considerations open the possibility that dynamics, usually associated with very high-energy phenomena may lie much closer to, and possibly within, accessible energies. If this were to be the case, a new puzzle arises: why would nature choose extremely small coupling constants? Since long ago~\cite{Dirac:1937ti,Dirac:1938mt} physicists have been reluctant to accept small (or large) numbers without an underlying dynamical explanation, 
even when the smallness of a parameter is technically natural in the sense of 't Hooft \cite{'tHooft:1979bh}. One reason for this reluctance is the belief that all physical quantities must eventually be calculable in a final theory with no free parameters. It would be strange for small numbers to pop up accidentally from the final theory without a reason that can be inferred from a low-energy perspective.  

In this work we propose a general mechanism to generate small numbers out of a theory with only ${\mathcal O}(1)$ parameters, and thus large effective interaction scales out of dynamics occurring at much lower energies. In all of these theories the full UV completion enters at energies exponentially smaller than suggested by a given interaction strength. The mechanism is fairly flexible and can produce exponentially large interaction scales for light or massless scalars, fermions, vectors, and even gravitons. It provides an interesting theoretical tool which opens new model-building avenues for axion, neutrino, flavour, weak scale, and gravitational physics. 

\begin{figure}[t]
\centering
\includegraphics[height=1.2in]{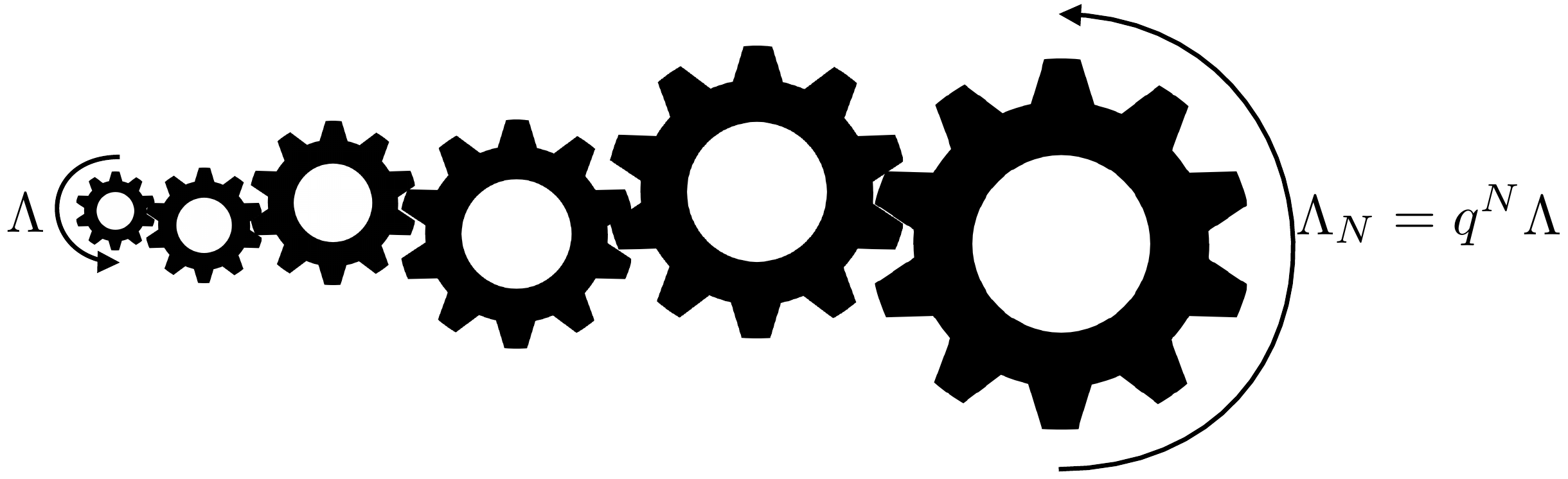}
\caption{A schematic representation of the clockwork mechanism increasing the interaction scale of a non-renormalisable operator.}
\label{fig:schem}
\end{figure}

The underlying structure is a generalisation of the clockwork models~\cite{Choi:2015fiu,Kaplan:2015fuy}, which were originally used to construct axion (or relaxion~\cite{Graham:2015cka}) setups in which the effective axion decay constant $f$ is much larger than the Planck mass $M_P$, without any explicit mass parameter in the fundamental theory exceeding $M_P$. In this way, one could circumvent the need for transplanckian field excursions in models which, for different phenomenological reasons, require $f>M_P$.  These constructions can be viewed as extensions of an original proposal for subplanckian completions of natural inflation~\cite{Kim:2004rp,Choi:2014rja,delaFuente:2014aca}. The name clockwork follows from the field phase rotations with periods that get successively larger from one field to the next (see \Fig{schem} for a pictorial interpretation).

The general framework is the following:  Consider a system involving a particle $P$, which remains massless because of a symmetry ${\cal S}$. At this stage neither the nature of $P$ or ${\cal S}$, nor whether the description is renormalisable or not, is crucial. We will give plenty of specific examples in our paper, but we want to stress that the general mechanism is insensitive to the details of the model implementation. 

Next, expand this simple setup to $N\! +\! 1$ copies of the original system, and consider them as sites of a one-dimensional lattice in theory space. The full theory possesses a symmetry ${\cal G}$, which contains at least the product of the individual symmetries (${\cal G}\supseteq {\cal S}^{N\! +\! 1}$), and describes $N\! +\! 1$ massless particles $P$. Now, at a mass scale $m$, introduce an explicit, but soft, breaking of the symmetry ${\cal G}$ which is local in theory space, through a mass mixing linking near-neighbours of the $N\! +\! 1$ lattice sites.  Thus far the construction resembles fields in a deconstructed flat extra dimension, \cite{ArkaniHamed:2001ca,Hill:2000mu,ArkaniHamed:2001nc}, however the critical difference is that this breaking includes a parameter $q\neq1$ that treats the site $j\! +\! 1$ and the site $j$ (with $j=0,..,N\! -\! 1$) \emph{asymmetrically}.  As we are considering a 1D lattice with boundaries, we can form only $N$ links out of the $N\! +\! 1$ sites. Since each link breaks the symmetry of a single site, one symmetry group ${\cal S}_0$ survives the breaking of ${\cal G}$; hence, one linear combination of the original particles (called $P_0$) remains massless. 

The crucial aspect of the clockwork is that the particle $P_0$ is not uniformly distributed in theory space along the sites, but is exponentially localised towards one of the boundaries. At the opposite boundary, the component of $P_0$ is exponentially small and is given by $1/q^N$. It is clear that, for moderately large values of $q$ and $N$ one can obtain exponential suppression of the $P_0$ component at one boundary. If, at that boundary, we couple the clockwork sector to the SM, we automatically obtain very small couplings of $P_0$ to ordinary particles, without introducing small parameters, multiple mass scales, or exponentially large field multiplicities in the fundamental theory.

When $P$ is a spin-0 scalar, spin-1/2 fermion, spin-1 boson, or spin-2 graviton, the corresponding symmetry ${\cal S}$ is a Goldstone shift symmetry, chiral symmetry, gauge symmetry, and linearised 4D diffeomorphism invariance, respectively. Each case introduces new model building applications. With a clockwork scalar one can construct invisible axion models at the weak scale. A clockwork fermion can explain a very small Dirac neutrino mass or address the hierarchical pattern of quark and lepton masses. With a clockwork gauge boson one can justify the existence of tiny gauge charges, with phenomenological predictions that differentiate from a scenario with simply a small gauge coupling. Finally, the most exciting application is found with the clockwork graviton, which offers a novel solution to the naturalness problem of the electroweak scale, providing a dynamical explanation for the weakness of gravity. In this paper, we will mainly focus on the structural aspects of the clockwork theory, leaving phenomenological issues to future work.

Importantly, the clockwork does not only have theoretical novelty, but comes together with experimentally testable consequences. Independently of the specific implementation of the clockwork, the theory predicts the existence of $N$ particles at the mass scale $m$, which will be called the `clockwork gears', as they are the degrees of freedom that make the mechanism work. As we will show, the gears have universal properties and their interactions with the SM can be predicted from the structural aspects of the clockwork implementation.

Our exploration of the clockwork will also lead us to consider the limit $N\! \to \! \infty$ and eventually the continuum limit, in which the 1-D lattice in field space is interpreted as a physical spatial dimension. This limit gives rise to non-trivial theories with finite clockworking effect. The simplest case has parameters $m$ and $q$ that are uniform along the extra dimension, wherein the geometry of this 5D space is uniquely defined. From this vantage point we build theories in a 5D clockwork space-time, which correspond to a continuum form of the clockwork theories for scalars, fermions, vectors, and gravitons.  Although the physical interpretation of the discrete and continuum clockworks are quite different, there is a simple correspondence between the two theories: the lattice in field space corresponds to the fifth spatial dimension; the clockwork gears correspond to the Kaluza-Klein modes; the interaction of an external sector with the last site of the discrete clockwork corresponds to the localisation of the external sector on a brane at the boundary of the compactified extra dimension.\footnote{This correspondence is familiar in deconstructions.}

Perhaps the most intriguing aspect of the clockwork is the link with the Higgs naturalness problem. The 5D interpretation elucidates how the clockwork can address this problem and its relation to previously proposed solutions, especially to Large Extra Dimensions (LED)~\cite{ArkaniHamed:1998rs} and  the warped extra dimensional model of Randall-Sundrum (RS)~\cite{Randall:1999ee}.  Remarkably, the metric of the clockwork space-time is identical to a 5D metric known as the linear dilaton model (see e.g.\ \cite{Antoniadis:2011qw}) which approximates the dual of Little String Theory (LST)\cite{Aharony:1998ub,Aharony:2004xn}, after compactifying additional dimensions (for more formal aspects see ~\cite{Dijkgraaf:1996hk,Dijkgraaf:1996cv,Berkooz:1997cq,Seiberg:1997zk,Dijkgraaf:1997nb,Losev:1997hx}). This may be a hint for an unexplored connection between string theory and the continuum version of the clockwork theory, as it identifies the clockwork solution of the hierarchy problem with the LST solution \cite{Lykken:1996fj,Antoniadis:1999rm,Antoniadis:2001sw,Antoniadis:2011qw}, including the corresponding phenomenological implications \cite{Antoniadis:1999rm,Antoniadis:2011qw,Baryakhtar:2012wj,Cox:2012ee}.

This paper is organised as follows.  In \Sec{secmechanism}, after discussing the difference between masses and interaction scales, we systematically construct the low energy effective theory of clockwork scalars, fermions, vectors and gravitons.  Throughout we will sketch phenomenological applications to axion models, light neutrino mass models, models of dark photons with millicharges, and multi-gravity theories.  The latter sets the scene for an explanation of the hierarchy between the Fermi and Planck scales.  In all of these models the number of fields is finite, and the theories are essentially simple 4D modules that may be straightforwardly employed for the aforementioned applications.  Following this, in \Sec{secdimension}, we go on to explore the continuum limit, where the number of fields is taken to infinity while the clockworking factor remains finite.  We find that in this limit the theory is best described as a 5D theory with a specific geometry.  A short summary of our results is given in \Sec{sec:conclusions}. Additional material is contained in five appendices.

\sectioneq{The clockwork mechanism}
\label{secmechanism}

\subsection{On masses, scales, and couplings}
\label{sec:massesscalescouplings}

The clockwork mechanism rests on the crucial difference between masses and interaction scales. Therefore, we find it useful to start by reviewing these concepts (a related discussion on the $\hbar$ expansion can be found in ref.~\cite{Brodsky:2010zk} and related discussions on derivative and field expansions are found in refs.~\cite{Manohar:1983md,Georgi:1989xy,Giudice:2007fh,Pomarol:2014dya,Panico:2015jxa}). 

To appreciate the difference between masses and scales it is useful to reinstate in the Lagrangian the appropriate powers of $\hbar$, while working in units with $c=1$. This means that time and length have identical units, while we distinguish between units of energy ($E$) and length ($L$). We start by considering a general 4D action involving scalar ($\phi$), fermion ($\psi$), and vector gauge fields ($A_\mu$), normalised such that all kinetic terms and commutation relations are canonical. Moreover, we express masses in units of inverse length, so that all mass parameters in the Lagrangian are written in terms of ${\tilde m} =m/\hbar$. In our basis, {\it there are no explicit factors of $\hbar$} in the classical Lagrangian in position space. With these assumptions, the dimensionality of the quantities of interest, including gauge couplings $g$, Yukawa couplings $y$, and scalar quartic couplings $\lambda$, are
\beq
[\hbar ] = EL ~,~~~[{\mathcal L}] =EL^{-3} ~,~~~ [\phi]=[A_\mu] = E^{1/2}L^{-1/2} ~,~~~ [\psi]=E^{1/2}L^{-1} ~,
\label{dimf}
\eeq
\beq
[\partial ] = [{\tilde m}]=L^{-1}~,~~~ [g]=[y]=E^{-1/2}L^{-1/2} ~,~~~[\lambda ] =E^{-1}L^{-1} ~.
\eeq
Canonical dimensions in natural units with $\hbar =1$ are recovered by identifying $E=L^{-1}$.

Note that $[g^2]=[y^2]=[\lambda]$, in agreement with the usual perturbative series. It is also important to remark that loop effects do not modify the dimensionality counting. Indeed, one can prove that, in our basis, each loop in momentum space carries one factor of $\hbar$. So each loop is accompanied by factors such as $\hbar g^2/(4 \pi)^2$, $\hbar y^2/(4 \pi)^2$, or $\hbar \lambda/(4 \pi)^2$, which are all dimensionless quantities in units of $L$ and $E$, and thus do not alter the dimensionality of the quantity under consideration.

Unlike the case of natural units, our dimensional analysis shows that couplings, and not only masses, are dimensionful quantities. Then, for our discussion, it is useful to introduce convenient units of mass ${\tilde M}\equiv L^{-1}$ and coupling $C\equiv  E^{-1/2}L^{-1/2}$.

Let us now add to the Lagrangian an effective operator of canonical dimension $d$ of the general form
\beq
\frac{1}{\Lambda^{d-4}}\, \partial^{n_D}\, \Phi^{n_B}\, \psi^{n_F} ~.
\eeq
Here $n_D$ is the number of derivatives, $n_B$ the number of boson fields ($\Phi =\phi ,\, A_\mu$), and $n_F$ the number of fermion fields, with $n_D+n_B + \frac 32 n_F =d$. The dimensionful quantity $\Lambda$ that defines the strength of the effective interaction will be called {\it scale}.
Its dimensionality is
\beq
[\Lambda ] = \frac{\tilde M}{C^{\frac{n-2}{d-4}}} ~,
\label{theeq}
\eeq
where $n=n_B+n_F$ is the total number of fields involved in the operator. This result can be immediately understood by recalling that each field carries an inverse power of $C$ ($[\phi]=[A_\mu]={\tilde M}C^{-1}$, $[\psi]={\tilde M}^{3/2}C^{-1}$) and the Lagrangian dimensionality is $[{\mathcal L}]={\tilde M}^4C^{-2}$.

Since the exponent of $C$ in \eq{theeq} is strictly positive, scales and masses are measured in different units and are not commensurable quantities. A scale is the ratio between a mass and a certain power of couplings. Equation~(\ref{theeq}) dictates the minimum number of couplings required to define the corresponding scale. If the operator is generated at the loop level in the fundamental theory, the number of couplings entering the scale $\Lambda$ can be larger than what \eq{theeq} prescribes. However, as previously discussed, these couplings are always accompanied by an appropriate power of $\hbar$ and do not alter the conclusion from dimensional analysis: {\it masses and scales are incommensurable}. 

Not only have masses and scales different dimensionality, but also carry different physical meanings. A mass is associated with $E_m={\tilde m} \hbar$, the energy at which new degrees of freedom appear. A scale is associated with 
$E_\Lambda = \Lambda \hbar^{\frac{2d-n-6}{2(d-4)}}$, the energy at which the theory becomes strongly coupled, if no new degrees of freedom intervene to modify the effective description. Therefore, a scale carries information on the strength of the interaction, but gives no information about the energy scale of new dynamics. The latter is given by the product of a scale times couplings, {\it i.e.} by mass.

To make the discussion more concrete, let us consider some examples of the relation between mass, scale, and coupling in familiar theories. The first example is the four-fermion interaction in the Fermi theory. Equation (\ref{theeq}) with $n=4$ and $d=6$ gives
\beq
[\Lambda] = [G_F^{-1/2}] = \frac{[M_W]}{[g]} ~.
\eeq
So $M_W$ is a mass and $G_F^{-1/2}$ a scale.
This is consistent with the notion that the new degrees of freedom in the electroweak theory occur at $E\sim M_W$ and not at $E \sim G_F^{-1/2}$. The latter is the energy scale at which perturbative unitarity would break down, in the absence of the weak gauge bosons. Note also that, since $G_F^{-1/2}\sim v$, the Higgs vacuum expectation value has the meaning of a scale, and not of a mass. Indeed, in the Higgs mechanism, physical masses are always given by the product of $v$ times a coupling constant. This result has a more general validity, which goes beyond the Higgs mechanism.
From \eq{dimf} we see that the vacuum expectation value of a scalar field has always the dimension of a scale $[\langle \phi \rangle ] ={\tilde M}/C$. For instance, the axion decay constant $f_a$ that defines the axion interactions in the low-energy effective Lagrangian is a scale and not a mass. Therefore, we cannot necessarily expect that the new physics associated with PQ breaking must occur at $E\sim f_a$, as will be confirmed by the clockwork.

Another example is the Weinberg operator $\ell \ell HH/\Lambda_\nu$ generating neutrino masses in the SM. In this case $n=4$ and $d=5$, and \eq{theeq} gives
\beq
[\Lambda_\nu]  = \frac{[M_R]}{[\lambda_\nu^2]} ~,
\eeq
where $M_R$ is the right-handed neutrino mass and $\lambda_\nu$ is the Yukawa coupling that participates in the see-saw mechanism. Since the physical neutrino mass is $m_\nu = v^2/\Lambda_\nu=\lambda_\nu^2 v^2/M_R$, we immediately see that the powers of couplings correctly match to give $m_\nu$ the dimension of mass. Thus the UV-completion of the Weinberg operator may enter at energies far below $\Lambda_\nu$.

Finally, let us consider the graviton coupling in linearised general relativity. With $n=3$ and $d=5$, \eq{theeq} gives
\beq
[M_P]  = \frac{[M_s]}{[g_s]} ~,
\eeq
where we can interpret $M_s$ and $g_s$ as the string mass and string coupling, respectively. From this perspective it is evident that $M_P$ is a scale and not a mass. Therefore, without any specific assumptions on the underlying couplings, we cannot conclude that the new degrees of freedom of quantum gravity must appear in the proximity of $M_P$.

These examples demonstrate that there is no technical obstruction, at least in field theory, to have extremely high energy interaction scales generated from a UV-completion which appears at much more pedestrian energies.  We will now see that the clockwork provides a concrete example where this separation of masses and interaction scales occurs without the introduction of any exponentially large or small parameters at the fundamental level.

\subsection{Clockwork scalar}
\label{sec:CWS}

The simplest way to implement the clockwork mechanism is with scalar fields~\cite{Choi:2015fiu,Kaplan:2015fuy}.  The implementations in \cite{Choi:2015fiu,Kaplan:2015fuy} involve renormalisable scalar field theories, however we will focus only on the low energy effective theory, which may have different UV-completions.  Let us consider a theory with a global symmetry ${\cal G}=U(1)^{N +1}$ spontaneously broken at the scale $f$. Below $f$, the effective degrees of freedom are $N\! +\! 1$ Goldstone bosons $\pi_j$, conveniently described in terms of the fields
\beq
U_j (x)= e^{i\pi_j (x)/f} ~~~~j=0, .. , N 
\eeq 
which transform by a phase under the corresponding Abelian factor $U(1)_j$. For simplicity, we assume that the spontaneous breaking of all Abelian factors contained in $\cal G$ occurs at the same scale $f$.

We also explicitly, but softly, break $\cal G$ by means of $N$ dimension-two parameters $m^2_j$ (with $j=0,\, ... , N\! -\! 1$), which can be regarded as the background values of $N$ spurion fields with charge 
\beq
Q_i [m^2_j ]=\delta_{ij} -q\, \delta_{i\,  j+1}
\label{charge_i}
\eeq
under the Abelian factor $U(1)_i$. We take $q>1$ and assume that the explicit breaking is small with respect to the scale of spontaneous breaking, {\it i.e.}  $m^2_j \ll f^2$. The smallness of $m^2_j / f^2$ is technically natural because $m^2_j / f^2\to 0$ enhances the symmetry of the theory. The hypothesis of a scale separation between $m^2_j$ and $f^2$ is the element that allows us to construct a low-energy effective theory of the pseudo-Goldstones $\pi_j$ from symmetry considerations alone, without committing to any specific UV completion at the scale $f$. 

The unbroken $U(1)$ corresponds to the generator 
\beq
{\cal Q} = \sum_{j=0}^N \frac{Q_j}{q^j} \, ,
\label{charge}
\eeq
where $Q_j$ are the generators of the Abelian factors in $\cal G$. Indeed, all of the parameters $m^2_j$ are neutral under the generator ${\cal Q}$, since \eq{charge_i} implies that ${\cal Q}[ m^2_j ]=0$ for any $j$. To simplify expressions, henceforth we take a single scale for the explicit breaking, {\it i.e.} $m^2_j \equiv m^2$. The generalisation to non-universal values of $f$ and $m^2$ for the different $U(1)$ factors is straightforward, and the physical content of the theory does not change, as long as we consider small deformations of the universal case. 

The low-energy description of the Goldstone boson and the $N$ pseudo-Goldstones is captured by an effective Lagrangian (formally $\cal G$-invariant, once we treat $m^2_j$ as spurions charged under $\cal G$), which can be expanded in derivatives and powers of $m^2$. The two leading terms are\footnote{Throughout the paper we use positive signature for the metric in flat space, $\eta =(-,+,+,+)$.}
\beq
{\mathcal L} = -\frac{f^2}{2}\, \sum_{j=0}^N \partial_\mu U_j^\dagger\, \partial^\mu U_j +\frac{m^2f^2}{2}\, \sum_{j=0}^{N-1} \left( U_j^\dagger\, U_{j+1}^q + \hc \right) \, .
\label{lag1}
\eeq
With no loss of generality the parameter $m^2$ can be chosen real (actually, even non-universal $m_j^2$ can all be made real simultaneously by an appropriate $\cal G$ transformation). 

In terms of the fields $\pi_j$, \eq{lag1} becomes
\beq
{\mathcal L} = -\frac{1}{2} \sum_{j=0}^N \partial_\mu \pi_j \partial^\mu \pi_j - V(\pi )
\label{lag2}
\eeq
\beq
V(\pi) =
\frac{m^2}{2} \sum_{j=0}^{N-1}
\left( \pi_j  -q\, \pi_{j+1}\right)^2 + {\mathcal O} (\pi^4)=\frac 12 \sum_{i,j=0}^N \pi_i \, M^2_{\pi \, ij}\, \pi_j + {\mathcal O} (\pi^4) \, .
\eeq
The mass matrix $M_\pi^2$ is given by
\beq
M^2_\pi = m^2
\begin{pmatrix}
1 & -q & 0 & \cdots &  & 0 \cr
-q & 1+q^2 & -q & \cdots &  & 0 \cr
0 & -q & 1+q^2 & \cdots & & 0 \cr
\vdots & \vdots & \vdots & \ddots & &\vdots \cr
 & & & & 1+q^2 & -q \cr
 0 & 0 & 0 &\cdots & -q & q^2
\end{pmatrix} \, .
\label{pimass}
 \eeq
The matrix $M_\pi^2$ becomes diagonal in the field basis $a_j$ ($j=0,\, ...,N$), related to the $\pi_j$ by a real $(N\! +\! 1)\! \times \! (N\! +\! 1)$ orthogonal matrix $O$ 
\beq
\pi = O\, a \, , ~~~~~~O^T M^2 O = {\rm diag}\, (m^2_{a_0},\dots , m^2_{a_N})
\eeq
where the eigenvalues are given by
\beq
m^2_{a_0} =0\, ,~~~m^2_{a_k} =\lambda_k\, m^2\, ,~~~~~ k=1, .. , N
\label{mass}
\eeq
\beq
\lambda_k \equiv q^2 + 1 -2q \cos \frac{k\pi}{N\! +\! 1} \, , ~~~ 
\label{lambdak}
\eeq
and the elements of the rotation matrix are 
\beq
O_{j 0} = \frac{{\cal N}_0}{q^j} \, , ~~~ 
O_{j k} = {\cal N}_k \left[ q \sin \frac{j k\pi}{N\! +\! 1}-  \sin \frac{(j +1) k\pi}{N\! +\! 1} \right] \, ,~~~ j =0, .. , N;~~k =1, .. , N
\label{rotation}
\eeq
\beq
{\cal N}_0 \equiv \sqrt{\frac{q^2-1}{q^2-q^{-2N}}} \, , ~~~~ {\cal N}_k \equiv \sqrt{\frac{2}{(N\! +\! 1)\lambda_k}} \, .
\eeq

Equation~(\ref{mass}) shows that the physical spectrum contains a massless Goldstone ($a_0$) and $N$ massive pseudo-Goldstone states   
($a_k$), which  are the ``gears" that make the clockwork mechanism work. The masses of the clockwork gears fill a band that ranges from $m_{a_1}\approx (q-1)m$ to $m_{a_N}\approx m_{a_1}+\Delta m$, with $\Delta m/m_{a_1}=2/(q-1)$. The mass splitting $\delta m_{k}=m_{a_{k+1}}-m_{a_k}$ between two successive states within the band $\Delta m$ is, in the large $N$ limit,
\beq
\frac{\delta m_{k}}{m_{a_k}}\approx \frac{q\pi}{N\lambda_k} \sin \frac{k\pi}{N\! +\! 1} \, ,~~~~~ k=1, .. , N\! -\! 1
\eeq
Although the exact expression in \eq{mass} is valid only for universal $f$ and $m^2$, non-universal deformations still preserve the structure of the clockwork gears: $N$ states within a mass band $\Delta m/m_a = {\mathcal O} (1)$ with splittings between successive states $\delta m/m_a = {\mathcal O} (1/N)$.

The crucial point of the clockwork lies in the expression of $O_{j 0}$ in \eq{rotation}. $O_{j 0}$ measures the component of the massless Goldstone boson contained in $\pi_j$. Since $O_{j 0}\propto q^{-j}$, the Goldstone component at each successive site is $q$ times smaller than for the previous site. This means that, for sufficiently large $N$, the Goldstone interaction can be very efficiently secluded away from the last site. If a theory is coupled to the clockwork sector only through its $N$-th site, the decay constant of the Goldstone interactions will appear exponentially enhanced with respect to the actual scale of spontaneous symmetry breaking.

To illustrate the mechanism, consider the case in which the $N$-th site $\pi_N$ is coupled to the topological term of a gauge theory
\beq
{\mathcal L} =\frac{ \pi_N}{16\pi^2 f} G_{\mu \nu} {\widetilde G}^{ \mu \nu} \, .
\label{gluon}
\eeq
Once we express $\pi_N$ in terms of mass eigenstates, using $\pi_N = \sum_{j=0}^N O_{Nj}\ a_j$, the effective interaction in \eq{gluon} becomes
\beq
{\mathcal L} =\frac{1}{16\pi^2}\, G_{\mu \nu} {\widetilde G}^{ \mu \nu}\left( \frac{a_0}{f_0}-\sum_{k=1}^N (-)^k\,\frac{a_k}{f_k} \right)
\label{gluon2}
\eeq
\beq
f_0 \equiv \frac{f q^N}{{\cal N}_0} \, , ~~~~ f_k \equiv \frac{f}{{\cal N}_k\ q \ \sin \frac{k\pi}{N\! +\! 1}} \, .
\label{fk}
\eeq

The first term in \eq{gluon2} exhibits the clockwork mechanism. The coupling of the Goldstone ($a_0$) to gauge bosons is determined by the effective scale $f_0$, which is exponentially enhanced with respect to the scale $f$ at which the symmetry-breaking dynamics takes place ($f_0/f \sim q^N$). 
From \eq{gluon2} we see that the clockwork gears inherit couplings to gauge bosons as well. However, their decay constants grow only mildly with $N$ ($f_k/f \sim N^{3/2}/k$) and are modulated by the index $k$. 

This mechanism allows for the construction of axion models in which the PQ-breaking dynamics can occur at or below scales as low as the weak scale, and yet the axion is nearly invisible. The model-building aspects, collider phenomenology, and cosmology of theories with weak-scale near-invisible axions are so rich and interesting that they will be presented in a separate publication.

\subsection{Clockwork fermion}
\label{discfermion}

Fermions may be kept massless due to a chiral symmetry, thus a fermion realisation of the clockwork involves a single chiral symmetry that is shared amongst a number of fields in the underlying model, such that the one remaining chiral symmetry pushes the massless fermion exponentially to one end of the clockwork.  To this end, let us introduce $N+1$ chiral fermions $\psi_{R \, j}$ ($j=0,..,N$) together with $N$ fermions $\psi_{L \, i}$ ($i=0,..,N-1$) of opposite chirality. Of course, the role of left and right chiralities can be reversed. The global chiral symmetry is broken by $N$ mass parameters $m_i$ that pair up the fields in $N$ massive Dirac fermions, leaving a single massless chiral component. The chiral symmetry is also broken by $N$ parameters $(mq)_i$ which can be regarded as the background values of spurion fields. It is useful to classify these parameters in terms of their charges under the global chiral symmetry. Let us call $U(1)_{R \, j}$ and $U(1)_{L \, i}$ the Abelian factors under which $\psi_{R \, j}$ and $\psi_{L \, i}$ have charge 1, respectively. Then $m_j$ have charges $(1,-1)$ under $U(1)_{L \, j}\times U(1)_{R\, j}$, and
$(mq)_j$ have charges $(1,-1)$ under $U(1)_{L \, j}\times U(1)_{R\, j\! + \! 1}$. One abelian factor of the chiral symmetry in the  $\psi_{R}$ sector is left unbroken by both $m$ and $mq$.

The Lagrangian for the fermion fields is 
\be
\mathcal{L} =  \mathcal{L}_{\rm Kin} - m  \sum_{j=0}^{N-1} \left( {\bar \psi}_{L \, j} \psi_{R \, j} -q\, {\bar \psi}_{L \, j} \psi_{R \, j\! + \! 1}
+ \hc \right) \equiv \mathcal{L}_{\rm Kin} - \left( {\bar \psi}_{L} M_\psi \psi_R + \hc \right) 
\label{lagfer}
\ee
where $\mathcal{L}_{\rm Kin}$ denotes the kinetic term for all fermions.
As in the scalar case, we take for simplicity universal values for $m$ and $q$. While $m$ can be chosen real with a chiral rotation of the fermions, the parameters $q$ are in general complex, but we will treat them as real, also for simplicity. Generalisations are straightforward.

The $N\times (N\! +\! 1)$ mass matrix $M_\psi$ is
\beq
M_\psi = m
\begin{pmatrix}
1 & -q & 0 & \cdots &  & 0 \cr
0 & 1 & -q & \cdots &  & 0 \cr
0 & 0 & 1 & \cdots & & 0 \cr
\vdots & \vdots & \vdots & \ddots & &\vdots \cr
 & & & & -q & 0 \cr
 0 & 0 & 0 &\cdots & 1 & -q
\end{pmatrix} \, .
\label{psimass}
 \eeq
The emergence of the clockwork mechanism is immediately clear, once we realise that $M_\psi^\dagger M_\psi$ is identical to $M_\pi^2$ in \eq{pimass}. 

The mass eigenstates $\Psi_L$ and $\Psi_R$ are given by
\beq
\psi_L = U_L \Psi_L \, , ~~~~ \psi_R = U_R \Psi_R
\eeq
\beq
U_R = O \, , ~~~~U_{L \, jk} = \sqrt{\frac{2}{N+1}} \sin \frac{jk\pi}{N+1} ~~~~j,k =1,\dots , N
\eeq
where the $(N\! +\! 1)\times (N\! +\! 1)$ matrix $O$ is defined in \eq{rotation}.

The spectrum consists of $N$ Dirac `fermion gears'  with masses
\beq
m_{\Psi_k} = m \sqrt{\lambda_k} ~~~~k =1,\dots , N
\eeq
where $\lambda_k$ are defined in \eq{lambdak}. The surviving chiral symmetry keeps one state massless, given by
\beq
\Psi_{R\, 0} ={\cal N}_0 \sum_{j=0}^N \frac{\psi_{R\, j}}{q^j} \, .
\eeq
As expected, the massless state has an overlap with $\psi_{R\, N}$, which is suppressed by a clockworking factor $q^N$.

The fermion clockwork has phenomenological applications to neutrino and flavour physics. The lightness of neutrinos is usually attributed to the see-saw mechanism, because having very small neutrino Yukawa couplings is viewed as a contrived possibility. However, the clockwork could give an explanation for a hierarchically small neutrino Yukawa.\footnote{Essentially, this would look like a discrete clockwork version of the higher-dimensional models of e.g. \cite{ArkaniHamed:1998vp}.}  Suppose that the theory preserves lepton number and that the Higgs ($H$) and left-handed lepton doublet ($L_L$) have a Yukawa coupling with the last site of a fermonic clockwork
\be
\mathcal{L} = - \lambda H {\bar L}_L \psi_{R \, N} + \hc 
\ee
The right-handed neutrino is identified with the light state ($\Psi_{R\, 0} $) left by the clockwork. Its effective Yukawa interaction is
\be
\mathcal{L} = - \lambda_0 H {\bar L}_L \Psi_{R \, 0} + \hc  ~~~~\lambda_0 =\lambda {\cal N}_0 q^{-N} 
\ee
A realistic neutrino mass can be obtained for $N \approx 25$ and $q\approx 3$, even if the original Yukawa coupling $\lambda$ is of order unity. 

Having a Dirac, rather then Majorana, neutrino mass would have significant impact on phenomenology predicting, in particular, that
neutrinoless double $\beta$-decay must be entirely absent. Moreover the mass scale of the fermion gears
could be accessible to experimental searches. Other phenomenological considerations could be made for the usual Yukawa couplings, assuming that the clockwork mechanism is responsible for the observed pattern of hierarchies.

\subsection{Clockwork photon}
A clockwork photon can be obtained by clockworking a gauge symmetry.  
Take $N\! +\! 1$ $\text{U}(1)$ gauge groups with equal gauge coupling $g$, and $N$ complex scalars 
$\phi_{j}$ ($j=0,..,N\! - \! 1$) each with charge $(1,-q)$ under the gauge groups $\text{U}(1)_{j}\times \text{U}(1)_{j\! +\! 1}$.  Give all of these scalars a negative mass-squared, which triggers vacuum expectation values which we assume to be at the same scale $f$.  The Lagrangian is
\be
\mathcal{L} = -\sum_{j=0}^{N} \frac{1}{4}F^j_{\mu\nu}  F^{j\, \mu\nu} - \sum_{j=0}^{N-1} \left[
|D_\mu \phi_j |^2 +\lambda (|\phi_j|^2 - f^2/2)^2\right]
\ee
\beq
D_\mu \phi_j \equiv \left[ \partial_\mu +ig \left( A_\mu^j -q A_\mu^{j\! + \! 1} \right) \right] \phi_j
\eeq 
The pattern of spontaneous symmetry breaking is $\text{U}(1)^{N+1} \to \text{U}(1)$.  Below the scale $f$, working in unitary gauge, we find the effective Lagrangian for the gauge fields
\be
\mathcal{L} = -\sum_{j=0}^{N} \frac{1}{4}F^j_{\mu\nu}  F^{j\, \mu\nu} + \sum_{j=0}^{N-1} 
 \frac{g^2f^2}{2}  (A^{j}_{\mu} - q A^{j+1}_\mu)^2 ~~.
\ee
The mass matrix is exactly of the clockwork form, with the heavy gauge bosons playing the role of the photon gear and one photon remaining massless.
 
The clockwork photon can have interesting phenomenological applications. If matter is charged only under the Abelian factor corresponding to the last site, the clockworking will generate exponentially small couplings to the massless photon. This can lead to visible particles with so-called millicharges, even though there are no small numbers in the theory.  Furthermore, heavy photon gears could be accessible to collider searches.  

\subsection{Clockwork graviton}
\label{sec:gravitonCW}

We conclude this section by turning our attention to the case of
 spin-$2$ gravitons.  
  
Let us imagine $N\! +\! 1$ copies of general relativity, with $N\! +\! 1$ associated massless gravitons.  In the linear approximation we can describe each graviton through an expansion of the metric around flat space-time, $g^{\mu\nu}_j = \eta^{\mu\nu}_j + 2\, h^{\mu\nu}_j/ M_j^2$.  The clockworking which breaks $N+1$ copies of diffeomorphism invariance to a single diffeomorphism invariance may be seen at the linear level through near-neighbour Pauli-Fierz terms for massive gravitons\footnote{Mass terms such as this typically arise in deconstructions of gravity \cite{ArkaniHamed:2002sp}.}
\be
\mathcal{L} = -\frac{m^2}{2} \, \sum_{j=0}^{N-1} \left( \left[ h_j^{\mu\nu} - q h_{j+1}^{\mu\nu} \right]^2 - \left[ \eta_{\mu\nu}  (h_j^{\mu\nu} - q h_{j+1}^{\mu\nu} ) \right]^2 \right)    ~~.
\label{pauli}
\ee

Limiting our considerations to the linear approximation, we see that the mass terms in \eq{pauli} are invariant under the gauge symmetry
\be
h_j^{\mu\nu} \to h_j^{\mu\nu} + \frac{1}{q^j} (\partial^\mu A^\nu +  \partial^\nu A^\mu) ~~,
\ee
where $A^\mu(x)$ is a space-time vector.  This gauge symmetry enforces the masslessness of the clockwork graviton and is respected by the clockwork structure of the mass terms.  The mass matrix is again of the clockwork form, with the heavy gravitons as the gears and one remaining massless graviton.  The massless graviton is described by the same linear combination of the original gauge eigenstates as for the scalar clockwork.

Suppose that the SM sector, with energy-momentum tensor $T^{\mu\nu}$, is coupled only to the last site of the clockwork, with a corresponding
`Planck-like' mass $M_N$. Then the coupling to the true massless graviton will be\footnote{Throughout the paper $M_P$ refers to the reduced Planck mass, equal to $2.4\times 10^{18}$~GeV.}
\be
-\frac{1}{M_N} h_N^{\mu\nu} T_{\mu\nu} \to  -\frac{1}{M_P} \tilde{h}_0^{\mu\nu} T_{\mu\nu}  ~~~~~~
M_P= \frac{q^N\, M_N}{{\cal N}_0}
~~.
\ee
We have found that the effective Planck scale $M_P$, which measures the strength of gravity in the low-energy sector of the theory, is exponentially larger than the fundamental gravity scale $M_N$, being enhanced by a factor $q^N$.
This offers the possibility of a clockwork solution to the hierarchy problem in which all new physics, including the completion of quantum gravity, may lie close to the weak scale, in full analogy with the solutions offered by LED or RS extra dimensions.

To understand how the clockwork can solve the hierarchy problem, we will explore an overarching framework which sheds new light on the clockwork. This is obtained by considering the limit in which the number of fields goes to infinity, $N\to \infty$, and the fields span a physical, albeit compactified, spatial dimension.\footnote{As discrete multi-gravity theories are plagued by theoretical subtleties we will focus on the continuum realisation of clockwork gravity; however it would be interesting to explore further the discrete theory sketched above.}

\sectioneq{A clockwork dimension}
\label{secdimension}
\subsection{The limit $N \! \to \! \infty$ and the approach to the continuum}

To develop a geometric picture of the clockwork mechanism, it is useful to consider the case in which
the discrete version of the clockwork arises as a deconstruction of an extra dimension.  Let us begin by defining an extra spatial coordinate $y$ with $-\pi R \leq y \leq \pi R$, where $R$ is the radius of the compactified dimension, orbifolded such that $y$ is identified with $-y$. 
We write the 5D metric in a reasonably general form
\be
ds^2 = X(|y|) dx^2 + Y(|y|) dy^2 ~~, ~~ dx^2 = -dt^2+d{\vec x}^2 ~~.
\label{eq:generalmetric}
\ee

The action for a real massless scalar in this space is\footnote{Throughout the paper we use a shorthand notation to indicate contraction of indices in flat space, $(\partial_\mu \phi )^2 \equiv \eta^{\mu \nu} \, \partial_\mu \phi \, \partial_\nu \phi$ with $\eta =(-,+,+,+)$.} 
\bea
{\mathcal S} & = &   2\int_0^{\pi R} dy\, \int d^4 x\, \sqrt{-g} \, \left(-\frac{1}{2} \, g^{MN}\, \partial_M \phi \, \partial_N \phi \right) \nonumber\\
& = & - \int_0^{\pi R} dy \, \int d^4 x\,  X^2 Y^{1/2}  \left[ \frac{(\partial_\mu \phi)^2}{X} + \frac{(\partial_y \phi)^2}{Y} \right] \nonumber\\
& = & - \int_0^{\pi R} dy \, \int d^4 x\,  \left[ (\partial_\mu \phi)^2 + \frac{X^2}{Y^{1/2}} \left( \partial_y \frac{\phi}{X^{1/2} Y^{1/4}}  \right)^2 \right] ~~,
\label{actionaction}
\eea
where $M,N$ are 5D space-time indices and in the last line a $y$-dependent field redefinition was made to realise canonical 4D kinetic terms. 

We now discretise the extra dimension by choosing $y_j = j a$ (with $j=0,..,N$) where $a$ is the lattice spacing, such that $N a= \pi R$. We also use the shorthand notation $F(j a)=F_j$ for $F=X,Y,\phi$.
After a trivial field rescaling, \Eq{actionaction} becomes
\be
{\mathcal S} =- \frac{1}{2} \int d^4 x \left[ \sum_{j=0}^N  (\partial_\mu \phi_j)^2 + \sum_{j=0}^{N-1} m_j^2 \left( \phi_j- q_j \phi_{j+1} \right)^2 \right] 
\ee
\beq
m_j^2 \equiv \frac{N^2\, X_j}{\pi^2 R^2\, Y_j} \, , ~~~q_j \equiv \frac{X_j^{1/2}Y_j^{1/4}}{X_{j+1}^{1/2}Y_{j+1}^{1/4}} ~.
\label{results}
\eeq

For the mass parameter $m_j^2$ to remain constant along the deconstruction, as in the clockwork, we must have $Y_j \propto X_j$.\footnote{Of course, as in the discrete models, it is not really important that the mass parameter is constant, but rather that it is at a similar scale along the lattice.  For simplicity we study only the constant case, but a generalisation to other cases would be interesting.} Furthermore, the only solution for $q$ to remain $y$-independent and for $q^N$ to give a finite but non-trivial clockworking in the limit of an infinite number of sites is\footnote{We are grateful to Riccardo Rattazzi for suggesting to us the use of this metric.}
\beq
X_j \propto Y_j \propto e^{-\frac{4k\pi R j }{3N}} \, ,
\label{discrete}
\eeq
such that
\beq
 q^N= e^{k \pi R} ~ .
\eeq
The parameter $k$, which will be called the `clockwork spring',  measures the effectiveness of the clockwork mechanism. When the clockwork is not operating (as in the case of a flat metric with $X=Y=1$), then $k =0$.

Therefore, in the large-$N$ limit of the discrete version, the clockwork parameters $m^2$ and $q$ must scale as
\beq
m^2 = \frac{N^2}{\pi^2 R^2} \, , ~~~~ q=e^{\frac{k\pi R}{N}} ~ .
\label{coarse}
\eeq

We can view \eq{coarse} as the Renormalisation Group (RG) trajectory of the clockwork parameters $m^2$ and $q$, as we coarse grain the extra dimension by changing $N$ for a fixed compactification radius $R$ or, equivalently, by changing the lattice spacing $a$. By defining the RG scale $\mu \equiv 1/a = N/\pi R$ and the $\beta$-functions as $\beta_X =d X/d\ln \mu$, we find 
\beq
\beta_{m^2} = 2 m^2 \, , ~~~~\beta_q =-q \ln q \, .
\eeq

It may seem from \eq{coarse} that the RG flow has an uninteresting behaviour in the UV, since both $m^2$ and $q$ have trivial UV attractors, $ m^2 \to \infty$ and $ q \to 1$ as $N\to \infty$. Instead, the UV limit of the discrete clockwork leads to a non-trivial theory. This can be seen by inspecting the mass spectrum as the clockwork parameters evolve according to their RG trajectory. Replacing \eq{coarse} in \eq{mass} and taking the large-$N$ limit, we find that the excitations are
\beq
m^2_0 =0 \, , ~~~ m_{n}^2 = k^2 + \frac{n^2}{R^2} + {\mathcal O}(1/N) ~~~~n=1, \dots , N
\label{masslarge}
\eeq
The gears have a characteristic spectrum, with evenly distributed energy levels and a mass-squared splitting equal to the inverse radius-squared. However, the spectrum is shifted by a mass gap equal to the clockwork spring $k$. The band $\Delta m$, which was finite at finite $N$, now extends to infinity as we take $N\to \infty$. 

\subsection{The continuum clockwork}

We now have the ingredients to study the clockwork from a 5D point of view. By extrapolating \eq{discrete} to the continuum, we find that the metric of the clockwork space is
\beq
ds^2 = e^{\frac{4 k |y|}{3}} (dx^2 + dy^2) \, .
\label{cmetric}
\eeq 
Note that we have flipped the sign of $k$. As discussed in \Appf{frames}, descriptions with positive or negative $k$ are equivalent and correspond to a change of coordinates. Our present choice is made to conform with phenomenological conventions in which the visible sector is located at $y=0$, rather than $y=\pi R$. As we will discuss in \sect{secUV}, the metric in \eq{cmetric} is the same as the one found in linear dilaton duals of LST~\cite{Antoniadis:2011qw}.

To allow for an easy interpolation between flat, warped, and clockwork spaces we rewrite the metric as 
\be
ds^2 = e^{\frac{4 k |y|}{3}} (dx^2 + e^{-4 \ell k|y|} dy^2) \, .
\label{eq:allmetric}
\ee
Flat space corresponds to $k=0$. For warped space, $\ell=1/3$ and $k=(3/2){\hat k}$ where ${\hat k}$ is the inverse AdS radius. We recover the conformally flat clockwork metric
of \eq{cmetric} using $\ell=0$.

The 5D action of a real massless scalar field in the geometry described by \eq{eq:allmetric} is
\be
{\mathcal S} = -\frac 12 \int d^4 x\, \int_{-\pi R}^{\pi R} dy \,  \left[ e^{2(1-\ell)k |y|}  (\partial_\mu \phi)^2 +  e^{2(1+\ell)k |y|} (\partial_y \phi)^2   \right]  ~~.
\label{uffaaction}
\ee
We expand the 5D field as
\beq
\phi(x,y) =\sum_{n=0}^{\infty} \frac{{\tilde \phi}_n(x)\, \psi_n(y)}{\sqrt{ \pi R}},
\label{phiexpand}
\eeq
where ${\tilde \phi}_n(x)$ satisfy the 4D free equation of motion $\partial_\mu^2 {\tilde \phi}_n(x) = m_n^2 {\tilde \phi}_n(x)$, while the equation of motion for $\psi_n(y)$ is
\be
\left[ \partial_y^2 -(1+\ell )^2{k}^2+ e^{-4\ell {k} |y|}m_n^2 \right] e^{(1+\ell ){k} |y|}\psi_n(|y|) =0 \, .
\label{eq:EOM}
\ee
For $\ell=1/3$, this is the usual equation for the KK modes in RS giving the mass eigenvalues $m_n =2k x_n /3= x_n {\hat k}$, where $x_n$ are the zeros of the Bessel function $J_1$.

Let us focus on the clockwork ($l=0$).  In this case \Eq{eq:EOM} becomes
\be
\left[ \partial_y^2 - k^2 +m_n^2 \right] e^{k |y|}\,
\psi_n(y) =0 \, .
\label{eq:EOM2}
\ee
Setting Neumann boundary conditions $\partial_y \psi=0$ at $y=0$ and $|y|=\pi R$ and normalising the modes $\psi$ such that ${\tilde \phi}$ have canonical kinetic terms in 4D, we find
\bea
\psi_0 (y) &=&  \sqrt{\frac{k \pi R}{ e^{2 k \pi R} -1 }}
\label{eigen1}\\
\psi_n (y) &=&  \frac{n}{m_n R} \, e^{-k |y|} \left(  \frac{k R}{n}\sin \frac{n |y|}{R}+\cos \frac{ny}{R} \right) ~~,~~~ n \in \mathbb{N}
\label{eigen2}
\eea
with mass
\beq
m_{0}^2  =  0  ~~,~~~~~~~~
m_n^2  =  k^2 + \frac{n^2}{R^2} ~~.
\label{uffamass}
\eeq

In the continuum, the gears play the role of the Kaluza-Klein (KK) excitations and their mass spectrum coincides with the result obtained in the large-$N$ limit of the discrete clockwork, see \eq{masslarge}.  From \eq{uffaaction} we see that, taking into account the integration measure, the density of the $n$-th KK mode is given by $dP= e^{2 k |y|}\, \psi_n^2(y)\, d(y/\pi R)$. Thus, the solutions in eqs.~(\ref{eigen1})--(\ref{eigen2}) show that the zero mode has a probability density exponentially localised at $y=\pi R$, while the excited modes have oscillating densities along the extra dimension. This is completely analogous to the case of the discrete clockwork, as exhibited by \eq{rotation}, once we recall that in the continuum case we have inverted the role of $y=0$ with $y=\pi R$ for phenomenological reasons.

The working of the continuum clockwork can be understood by considering an axion model where a 5D complex scalar spontaneously breaks a global $\text{U}(1)$ symmetry.  On a brane living at $y=0$ one can add a gauge group and matter fermions, together with a local interaction to the bulk scalar.  At low energies the effective theory for the axion $\phi$ is described by the action 
\be
{\mathcal S} =  \int d^4 x\, \int_{-\pi R}^{\pi R} dy \left[ -e^{2 k |y|}\, \frac{(\partial_M \phi )^2}{2} + 
 \delta(y) \left(-\frac{1}{4g^2} G_{\mu\nu} G^{\mu\nu} + \frac{\sqrt{\pi R}}{16 \pi^2 f}\, \phi \, G_{\mu\nu} \widetilde{G}^{\mu\nu} \right) \right] ~~,
 \label{actionS}
\ee
where we have included the covariant $\delta$-function, $\delta(y)/\sqrt{g_{55}}$, and 4D Levi-Civita symbol, $\epsilon^{\mu \nu \rho \sigma}/\sqrt {-g^{(4D)}}$, and
all index contractions are performed with a flat metric, as the clockwork factors have been explicitly extracted. The theory described by \eq{actionS} is exactly the analogue of \eq{gluon} for the discrete case, with the coupling to the last site of the discrete clockwork replaced by the coupling to a brane at the origin of the clockwork dimension. 

After expanding $\phi (x,y)$ as in \eq{phiexpand} and integrating over the extra dimension, \eq{actionS} becomes
\be
{\mathcal S} = \int d^4x \left(  -\frac{1}{2}\sum_n \left[ 
  ( \partial_\mu \tilde{\phi}_n)^2 + m_n^2 \tilde{\phi}_n^2 \right] 
   -\frac{1}{4g^2} G_{\mu\nu} G^{\mu\nu} +\frac{1}{16\pi^2}G_{\mu\nu} {\widetilde G}^{\mu\nu} \sum_{n= 0}^\infty  \frac{\tilde{\phi}_n}{f_n} 
   \right) ~,
\ee
where $f_n = f/\psi_n(0)$. Using eqs.~(\ref{eigen1})--(\ref{eigen2}), we obtain
\beq
\frac{f_0}{f} \approx \frac{e^{k\pi R}}{\sqrt{k\pi R}} ~,~~~~~~~~\frac{f_n}{f} =\sqrt{1+\frac{k^2R^2}{n^2}} ~,~~n=1,2,\dots 
\label{great}
\eeq
The decay constant for the zero mode is exponentially amplified with respect to the original $f$, just as in the discrete case, with $q^N$ replaced by its counterpart in the continuum $e^{k \pi R}$. On the other hand, the decay constants for the $n$-th excited gear remains roughly equal to $f$.

This demonstrates the continuum limit of the clockwork mechanism.  Let us now investigate how such a 5D set up can arise self-consistently.

\subsection{A clockwork geometry}
\label{secgeometry}

As discussed in ref.~\cite{Cox:2012ee}, the simplest setup that generates the desired metric is given by a dilaton field in 5D space-time. Let us define the theory in terms of the 5D gravity action in the Jordan frame 
\beq
{\mathcal S} = \int d^4 x\, dy\, \sqrt{-g} \, \frac{M_5^3}{2} e^{S} \left( {\mathcal R} +   g^{MN}\partial_M S
\, \partial_N S +4k^2 \right) ~,
\label{actionstring}
\eeq
where $S$ is the dimensionless dilaton field and $k^2$ characterises the (negative) vacuum energy in the bulk. The reason for our  normalisation of the vacuum energy term will be clear soon.  It should be kept in mind that this  $k^2$ term breaks a symmetry under which $S$ is shifted by a constant ($S\to S+c$) and the metric is rescaled by a constant Weyl factor ($g_{MN}\to e^{-2c/3}\, g_{MN}$).

We compactify the fifth dimension on an $S_1/Z_2$ orbifold with extra fields localised on its fixed points $y_0=0$ and $y_\pi=\pi R$. Calling $\Lambda_0$ and $\Lambda_\pi$ the corresponding vacuum energies, we add to the action the brane terms
\beq
{\mathcal S} = \int d^4 x\, dy\, \sqrt{-g} \, e^{S} \left[ -\frac{\delta (y -y_0)}{\sqrt{g_{55}}}\Lambda_0 -
\frac{\delta (y -y_\pi)}{\sqrt{g_{55}}}\Lambda_\pi \right] ~.
\eeq

It is convenient to work in the Einstein frame, where the gravity kinetic term is canonical. This is achieved through the metric transformation
\beq
g_{MN} \to e^{-\frac{2S}{3}} g_{MN}
\label{tram}
\eeq
which turns the total action into
\bea
{\mathcal S} &=& \int d^4 x\, dy\, \sqrt{-g}  \left\{ \frac{M_5^3}{2} \left( {\mathcal R} -\frac{1}{3} \,  g^{MN}\partial_M S
\, \partial_N S +e^{-\frac{2S}{3}}\, 4k^2 \right)   \right. \nonumber \\ 
& -&\left.  \frac{e^{-\frac{S}{3}}}{\sqrt{g_{55}}}\, 
 \left[\delta (y -y_0)\Lambda_0 +
\delta (y -y_\pi)\Lambda_\pi \right] \right\}
~.
\eea
Note that the canonically normalised dilaton field is $M_5^{3/2}S/\sqrt{3}$. In the Einstein frame it is apparent how the bulk action has a shift symmetry in $S$ in the limit $k \to 0$. This is important because, for phenomenological reasons, we are also interested in the case $k \ll M_5$. The shift symmetry ensures that this condition is technically natural and protected against quantum corrections.

We solve the  equations of motion for the theory assuming that the metric is consistent with Poincar\'e invariance in 4D and we fix the parametrisation invariance of the fifth coordinate by going to a conformally flat basis
\beq
ds^2 = e^{2\sigma (y)} \left( \eta_{\mu \nu} dx^\mu dx^\nu + dy^2 \right)~.
\label{ansatz}
\eeq
With this choice, the coupled system of differential equations for the $55$ and $\mu\nu$ components of the Einstein equations together with the equation of motion for $S$ is (see \Appf{5D} for a derivation)
\beq
\left\{ \begin{array}{l}
36\, \sigma'^2 -S'^2=12 \, k^2\, e^{2\left( \sigma - \frac{S}{3}  \right)} \\
9( \sigma''-\sigma'^2 )+ S'^2=-3\Delta \\
S''+3\sigma' S' =4\, k^2 \, e^{2(\sigma -\frac{S}{3}  )}- \Delta
\end{array}
\right.
\label{system}
\eeq
where primes denote derivatives with respect to the fifth coordinate $y$ and the boundary term $\Delta$ is
\beq
\Delta = \frac{e^{\left( \sigma -\frac{S}{3} \right)}}{M_5^3}\left[ \delta(y-y_0)\Lambda_0 +\delta(y-y_\pi)\Lambda_\pi \right] ~.
\eeq

Using the technique of ref.~\cite{DeWolfe:1999cp}, one can show that the most general solution of the system (\ref{system}), consistent with the four junction conditions on the derivatives of $\sigma$ and $S$ dictated by $\Delta$ and with the orbifold symmetry $y \to -y$, is
\beq
\sigma = \frac{2 k |y|}{3} \, e^{\left( \sigma_0 -\frac{S_0}{3} \right) } +\sigma_0 ~,~~~~
S = 2 k |y| \, e^{\left( \sigma_0 -\frac{S_0}{3} \right) }+S_0 ~,
\eeq
under the special conditions
\beq
-\Lambda_0 = \Lambda_\pi = 4 k M_5^3 ~.
\eeq
The two integration constants $\sigma_0$ and $S_0$ have no physical consequence.  Without loss of generality, we can choose $\sigma_0 =S_0=0$, so that the solution is simply 
\beq
3\sigma = S = 2k|y| ~.
\label{linear}
\eeq
We recognise that this solution indeed corresponds to the metric in \eq{cmetric}, derived from an extrapolation to the continuum of the discrete clockwork. In this context, the clockwork spring $k$ is interpreted as a measure of the bulk Jordan-frame vacuum energy of the compactified space in which the dilaton and gravity live.

Recalling from \eq{tram} that the relation between the Jordan and Einsten frame metrics is $g_{MN}^{(J)} =e^{-2S/3}g_{MN}^{(E)}$, the ansatz in \eq{ansatz} corresponds to
\beq
g_{MN}^{(J)} =e^{2\left( \sigma -\frac{S}{3} \right)} \eta_{MN} ~.
\eeq
Therefore, on the solution of \eq{linear}, space-time is flat in the Jordan frame and its intrinsic curvature vanishes. However, in this frame, the effective Planck mass is exponentially decreasing as we move along the fifth dimension towards $y\to 0$, signalling that gravity becomes prematurely strongly interacting near the $y=0$ brane. On the contrary, the Planck mass is constant in the Einstein frame, but the curvature grows exponentially as $y\to 0$ (see \eq{curvatureCW} in \Appf{5D}), revealing the onset of strongly-interacting gravity.

As noted in ref.~\cite{Cox:2012ee}, an appealing aspect of the theory defined by \eq{actionstring} is that a mechanism for radius stabilisation is already built in and does not require any additional field, unlike the RS case in which radius stabilisation is achieved at the expense of at least one new scalar field, as in the Goldberger-Wise solution~\cite{Goldberger:1999uk}. Indeed, let us suppose that interactions localised on the brane at $y=\pi R$ generate a potential for $S$, which fix the field value on the brane. This corresponds to an additional boundary condition $S(\pi R) =S_\pi$, where $S_\pi$ is a number naturally expected to be of order one. Imposing this boundary condition on the solution in \eq{linear} determines the value of $R$ such that
\beq
k\pi R =\frac{S_\pi}{2} ~.
\eeq
An efficient clockworking factor can be easily obtained for values of $S_\pi$ that are moderately large, but not incompatible with natural expectations.

Having the same field -- the dilaton -- responsible for both generating the non-trivial metric and stabilising the size of the compactified dimension is certainly an attractive feature of the theory. In the RS case, the geometry is determined by the vacuum terms, while an additional scalar field determines the brane separation. The counting of the required degrees of freedom is the same in both theories. We also remark that, while the clockwork has a radius stabilisation mechanism already built in, it is nonetheless compatible with solutions \`a la Goldberger-Wise, if a boundary condition on $S(\pi R)$ is not imposed. An example is given in \Appf{GW}.

\subsection{A solution to the hierarchy problem}
\label{sec:hierarchy}
On the background of the clockwork metric, the graviton fluctuations around 4D Minkowski space, in the transverse-traceless gauge and weak-field limit, are described by the action (see \Appf{gravaction} for a derivation)
\be
S   =  -\frac{1}{2}  \int d^4x \int_{-\pi R}^{\pi R} dy\, e^{2 k |y|} \bigg[ ( \partial_{\lambda} h_{\mu\nu}) ( \partial^{\lambda} h^{\mu\nu})+ (\partial_y h_{\mu\nu}) (\partial_y h^{\mu\nu}) \bigg] ~~.
\label{eq:finalaction}
\ee
This is the same form of action as for the scalar in \eq{uffaaction}.  Thus, decomposing the graviton mass eigenstates as
\beq
h_{\mu \nu} (x,y) =\sum_{n=0}^{\infty} \frac{{\tilde h}^{(n)}_{\mu \nu}(x)\, \psi_n(y)}{\sqrt{ \pi R}} ~ ,
\eeq
it is easy to see that the functions $\psi_n$ and the mass eigenvalues $m_n$ are given by eqs.~(\ref{eigen1})--(\ref{uffamass}), {\it i.e.} by the same solutions as in the scalar case.

Suppose that the SM sector with Lagrangian density ${\mathcal L}^{SM}(x)$ is localised on a 4D brane at $y=0$. Taking into account that the modes ${\tilde h}^{(n)}_{\mu \nu}$ have canonical kinetic terms in 4D, we can write the gravitational interaction as
\beq
\mathcal{L} = -\frac{h_{\mu \nu} (x, 0)\, T^{SM}_{\mu\nu}(x)}{M_5^{3/2}}=-\sum_{n=0}^\infty \frac{{\tilde h}^{(n)}_{\mu \nu} (x)\, T^{SM}_{\mu\nu}(x)}{\Lambda_n}
\eeq
\beq
T^{SM}_{\mu\nu}=\left. -2\frac{\partial {\mathcal L}^{SM}}{\partial g^{\mu \nu}}+ g_{\mu \nu}{\mathcal L}^{SM}\right|_{g_{\mu \nu} =\eta_{\mu \nu}}
~,~~~~
\Lambda_n \equiv \frac{\sqrt{\pi R} \, M_5^{3/2}}{\psi_n(0)} ~.
\label{efflambda}
\eeq

It is now useful to derive the effective 4D Planck mass $M_P$, defined as the constant in front of the Einstein-Hilbert action (see \Appf{5D}),
\beq
M_P^2 =2M_5^3 \int_0^{\pi R} dy \, e^{2ky}=\frac{M_5^3}{k}\left( e^{2k\pi R} -1 \right) ~.
\label{planckplanck}
\eeq
It is apparent how the clockwork can produce an effective 4D Planck mass exponentially larger than the fundamental 5D mass $M_5$.

Rewriting \eq{efflambda} with the help of \eq{planckplanck} and  the expressions of  $\psi_n$ in eqs.~(\ref{eigen1})--(\ref{eigen2}), we find that the effective scales of gravitational interaction are
\beq
\Lambda_0 = M_P ~,~~~~~~~~\Lambda_n = \sqrt{ M_5^3\, \pi R \left( 1 + \frac{k^2R^2}{n^2} \right)} ~.
\label{gravscales}
\eeq

Equation~(\ref{gravscales}) is the expression of the clockwork. It shows that the strength of the gravitational interaction of the massless graviton is determined exactly by the conventional Planck mass $M_P$. Instead, the interaction scale of the massive graviton gears is roughly given by $\Lambda_n \approx M_5^{3/2}/k^{1/2}$, which is smaller than $M_P$ by a clockworking factor $e^{k\pi R}$.

It is important to remark that $\Lambda_n$ measures the interaction of the graviton gears, but does not correspond to the scale at which perturbation theory ceases to be valid. Indeed,
the clockwork theory becomes strongly interacting at a scale much lower than $\Lambda_n$. This can be understood with the following argument.

Let us parametrise the production cross section of a single graviton gear as $\sigma_n =c /(\pi \Lambda_n^2)$. Here $c$ is a coefficient that depends on the production process under consideration, but does not depend on any of the clockwork parameters, as long as the energy involved $E$ is much larger than $m_n$. When $m_n$ becomes of the order of $E$ or larger, then $c$ quickly drops to zero. This means that, for a given energy $E$, one can produce only modes with $n\lsim N_{\rm max}$, where $N_{\rm max}=R(E^2-k^2)^{1/2}$.
The total cross section inclusive of all allowed channels is
\beq
\sigma = \sum_{n=0}^{N_{\rm max}}\sigma_n  \approx \frac{c\, E}{\pi^2 M_5^3} ~,
\eeq
where we have taken $E\gg k$. The condition of perturbative unitarity $\sigma \lsim \pi / E^2$ implies $E\lsim (\pi /c^{1/3})M_5$. This shows that $M_5$ is the scale at which the theory becomes strongly interacting and quantum gravity effects take over. The same conclusion could have been reached by considering scattering processes in 5D, as $M_5$ is evidently the effective scale of gravitational interactions.

We can also learn about the onset of strong dynamics with the following line of reasoning.
When the decay width of the graviton gears ($\Gamma_n$) is of the order of the mass splitting ($\delta m_n$), we lose the notion of separate particle excitations and perturbation theory breaks down. Since $\Gamma_n \approx N_{\rm ch} m_n^3/(16\pi \Lambda_n^2)$, where $N_{\rm ch}$ is the effective number of decay channels, and $\delta m_n\approx n m_n/(kR)^2$, a calculable perturbative regime requires a certain amount of separation between the gear mass and interaction scale, {\it i.e.} $m^2_n < n\Lambda^2_n$ (as we are working at fixed clockworking factor with $kR$ of order unity). At small and moderate values of $n$, this constraint implies $k<M_5$.

A final consideration arises from the fact that a sensible derivative expansion of the gravitational action is only possible whenever the curvature does not exceed the 5D Planck mass. This is a non-trivial constraint for the clockwork space, since its
 curvature is not constant. From the expression of the clockwork curvature, see \eq{curvatureCW}, we find that 
 the upper bound  $|\mathcal{R}| < M_5^2$ implies $k<M_5$, which is the same condition for perturbativity we have just derived above. This means that, in the regime in which the graviton gears are weakly interacting, the curvature condition is automatically satisfied.

The results presented in this section illustrate how the clockwork can solve the Higgs naturalness problem. Since $M_5$ is the cutoff scale of the theory, where quantum gravity takes over, the Higgs mass is naturally expected to be of the order of the fundamental scale $M_5$. Then the naturalness problem is solved by assuming that $M_5$ lies around the weak scale, while the 4D Planck mass is clockworked away to much larger values. In order to achieve this, the product $kR$ need only be moderately large. From \eq{planckplanck} we obtain
\beq
kR = 10 + \frac{1}{2\pi} \ln \left( \frac{k}{\rm TeV}\right) -\frac{3}{2\pi} \ln \left( \frac{M_5}{10~{\rm TeV}} \right)~.
\label{eccokR}
\eeq   
Since the effects of quantum gravity have not shown up at the LHC some degree of tuning is required to have $m_h < M_5$. However this is similar to the tuning required in other symmetry-based solutions to the hierarchy problem.

In conclusion, clockwork gravity in the continuum offers a solution to the Higgs naturalness problem, concluding this speculation from the discrete perspective of \Sec{sec:gravitonCW}.  The hierarchy $G_N/G_F$ is explained by taking a theory which must be UV-completed near the weak scale and clockworking the interaction scale of the massless graviton to $M_P$.  As in the scalar clockwork, the masses of the new resonances are not around $M_P$, since this very large interaction scale is just a mirage constructed from a weak-scale quantity and an exponentially small number.  As we will see, this is in fact the same solution as proposed in \cite{Antoniadis:2001sw,Antoniadis:2011qw} from the perspective of Little String Theory.

\subsection{UV perspective}
\label{secUV}

In string theory the 4D Planck mass can be related to the string coupling $g_s$, the string scale $M_s$, and the volume of the six extra dimensions $V_6$ as
\be
M_P^2 = \frac{M_s^8\, V_6}{g_s^2}  ~~.
\label{eq:LSTrel}
\ee
If $M_s \sim V_6^{-1/6} \sim $ TeV then a large 4D Planck mass arises in the limit of vanishingly small string coupling, $g_s \sim 10^{-15}(M_s/\text{TeV})(M_s^6 V_6)^{1/2}$.  This limit corresponds to a class of scenarios known as Little String Theory (LST)~\cite{Losev:1997hx,Berkooz:1997cq,Seiberg:1997zk,Dijkgraaf:1997nb,Dijkgraaf:1996hk,Dijkgraaf:1996cv}.  It was realised some time ago that in these theories the hierarchy problem may be resolved by bringing the string and compactification scales all the way down to the weak scale, and translating the puzzling ratio $G_N/G_F$ to a question of the smallness of the string coupling \cite{Lykken:1996fj,Antoniadis:1999rm,Antoniadis:2001sw,Antoniadis:2011qw}.  This approach is reminiscent of the considerations made in \Sec{sec:intro}, where it was argued that the UV-completion of gravity could lie far below the 4D Planck mass if a coupling is very small.

Using the AdS/CFT correspondence~\cite{Maldacena:1997re}, it has been argued that LST is described holographically by asymptotically linear dilaton backgrounds \cite{Aharony:1998ub,Aharony:2004xn}.  This has allowed the construction of calculable setups which address the hierarchy problem~\cite{Antoniadis:2001sw,Antoniadis:2011qw}. In a similar fashion, these considerations bring us to the linear dilaton background of \Eq{linear}, which  sources the clockwork metric and was derived from the continuum limit of the low energy description of the clockwork mechanism.  Thus a connection between the clockwork mechanism and LST is established.  

In retrospect, the connection between the two theories is not surprising.
In the clockwork mechanism an apparently large interaction scale $f_N$ is generated from a smaller scale $f$ through the introduction of a naturally small factor $q^{-N}$ (in other words $f_N \approx f/q^{-N}$).  In LST an exponentially large Planck scale $M_P$ is generated from TeV-size masses ($M_s$, $V_6^{-1/6}$) through the introduction of a small coupling $g_s$, as demonstrated in \Eq{eq:LSTrel}.  Thus the two made use of the same ingredients from the outset.  Furthermore, this suggests that the discrete clockwork gravity model of \Sec{sec:gravitonCW} is a deconstruction of the holographic dual of LST.  It should be kept in mind that this is not a deconstruction of LST itself, as the deconstruction of LST will be supersymmetric and conformal.  The explicit deconstruction of the full 6D $\mathcal{N}=(1,1)$ LST has been provided in \cite{ArkaniHamed:2001ie}.

Studies of LST also shed light on the possibility of a 4D field theory dual of the 5D clockwork theory.  Stacks of NS5-branes give rise to 6D LSTs which are strongly coupled, non-local, and do not admit a Lagrangian description \cite{Antoniadis:1999rm,Aharony:1998ub}.  Since the holographic dual of these theories is a linear dilaton background in higher dimensions, this would conversely suggest that a 4D theory that gives rise to a spectrum of 4D particles with the characteristic clockwork spectrum and interactions may correspond to a theory with very unusual properties, from a field theoretical point of view.  It is worth noting that the continuum clockwork theory is not AdS$_5$, thus it is not clear if a 4D dual theory exists.

Finally we stress that LST was recognised not only to offer a rationale for the hierarchy problem, but also as a source of potentially rich weak-scale phenomenology~\cite{Antoniadis:1999rm,Antoniadis:2011qw,Baryakhtar:2012wj,Cox:2012ee}, due to its peculiar particle spectrum.  The phenomenological studies were made possible through the dual linear dilaton theory description and, as the setup is the same, these studies also apply to the continuum clockwork theory.  We will return to the phenomenological consequences of the clockwork in a forthcoming publication.

\subsection{Relation to other theories}
The relationship between the clockwork gravity theory and LED or RS can also shed some light on the nature of the solution to the hierarchy problem.   

Let us first compare the discrete clockwork with deconstructions of LED, corresponding to $X_j=Y_j=1$ in \Eq{eq:generalmetric}, and to RS, which correspond to $X_j = \exp (-2{\hat k} \pi Rj/N)$, $Y_j=1$, where $\hat k$ is the inverse AdS radius. Using \eq{results}, we find that the parameters $m_j^2$ and $q_j$ that characterise the deconstruction are
\beq
\begin{array}{cc|cccc} 
&&& m_j^2 &~& q_j  \\[0.2cm]
 \hline \\[-0.3cm]
 {\rm LED} &&&\dfrac{N^2}{\pi^2 R^2}  &~& 1 \\[0.4cm]
 {\rm RS} &&&\dfrac{N^2}{\pi^2 R^2}e^{-\frac{2{\hat k}\pi Rj}{N}}   &~& e^{\frac{{\hat k}\pi R}{N}} \\[0.4cm]
  {\rm CW} &&&\dfrac{N^2}{\pi^2 R^2}  &~& e^{\frac{k \pi R}{N}}  
\end{array}
\eeq
For the LED case, the mass terms are site-independent, but there is no clockworking, as $q_j=1$. For RS models, the mass terms are warped along the extra dimension, thus they are site-dependent and do not all enter at the same scale.  However, the mixing term $q_j$, and hence the zero mode, is analogous to the clockwork.  This suggests that the clockwork metric is rather unique.  It realises site-independent mass terms which all enter at the same mass scale, as in LED, but with warping of bulk zero mode interactions, as in RS models.

These differences are also found in the continuum version of the theories. The masses of the KK modes, their interaction scale, and the relationship with the 4D Planck mass in the three 5D theories are given by
\beq
\begin{array}{cc|cccccc} 
 &&& m_n^2 &~& \Lambda_n^2 &~& M_P^2 \\[0.2cm]
 \hline \\[-0.3cm]
 {\rm LED} &&& \dfrac{n^2}{R^2} &~& \dfrac{M_P^2}{2} &~& M_5^3\, 2\pi R \\[0.4cm]
 {\rm RS} &&& \approx [ ( n+\frac 14 ) \pi {\hat k} ]^2 &~&  \approx \dfrac{M_5^3}{\hat k} &~&  \dfrac{M_5^3}{\hat k} (e^{2{\hat k}\pi R} -1) \\[0.4cm]
 {\rm CK} &&& k^2 + \dfrac{n^2}{R^2} &~& M_5^3 \pi R \left(1+\frac{k^2R^2}{n^2}\right) &~&  \dfrac{M_5^3}{k} (e^{2{k}\pi R} -1)
\end{array}
\label{tabella}
\eeq 

In LED, the ratio between the Planck and weak scale is explained by a large volume factor ($V\gg M_5^{-1}$, where the extra dimensional `volume' is $V=2\pi R$). The KK modes have $M_P$-suppressed interactions\footnote{The factor of $1/2$ in $\Lambda_n^2$ for LED comes because we are restricting the summation over KK modes to positive values of $n$. The KK modes with negative $n$ are identified with their $-n$ counterpart and the fields are finally rescaled to make the kinetic terms canonical.}
and are uniformly distributed in mass. The large volume implies very small KK mass splittings, $\delta m_n =2\pi M_5^3/M_P^2$.

In RS, one generally assumes $M_5 \sim {\hat k}$. Then the hierarchy is explained not by the volume but by the geometry, with a warping factor such that $M_P \sim M_5 e^{{\hat k}\pi R}$. The KK modes have TeV-scale interactions and masses characterised by $\hat k$, with an approximately uniform distribution such that $\delta m_n/m_n = {\mathcal O} (1)$ (for instance, $(m_2-m_1)/m_1 = 0.84$). The RS expressions for $m_n$ and $\Lambda_n$ in \eq{tabella} are valid only in the limit of large warping ($e^{{\hat k}\pi R}\gg 1$). Using instead expressions valid in the limit of small $\hat k$, one can show that RS coincides with LED when  ${\hat k} \to 0$. Thus, RS can be viewed as a deformation of LED controlled by the parameter ${\hat k}$. This result was used in ref.~\cite{Giudice:2004mg} to construct an RS theory with small AdS curvature (${\hat k} \ll M_5$). In the UV the theory is identical to 5D LED, while the small warping creates a mass gap in the IR. In this way, astrophysical bounds on light KK emission are avoided, while the collider phenomenology of 5D LED (often unjustly neglected by experimental analyses) is viable. Modified KK spectra have also been considered in the context of hyperbolic geometries, such as \cite{Kaloper:2000jb}.

The clockwork can be viewed as an alternative deformation of 5D LED. In the limit $k\to 0$, the clockwork expressions in \eq{tabella} coincide with those of LED. However, for finite $k$, there are important differences. The mass spectrum has a mass gap, followed by a dense distribution of states such that, for moderate values of $n$,
\beq
\frac{\delta m_n}{m_n} \approx \frac{n+\frac 12}{k^2R^2} ~.
\eeq
Using \eq{eccokR}, we see that for $m_1$ in the TeV range, the mass splitting is in the tens of GeV and, depending on the energy resolution and smearing, at collider experiments the resonances can be individually identified or appear as a continuum.

\begin{figure}[t]
\centering
\includegraphics[height=1.4in]{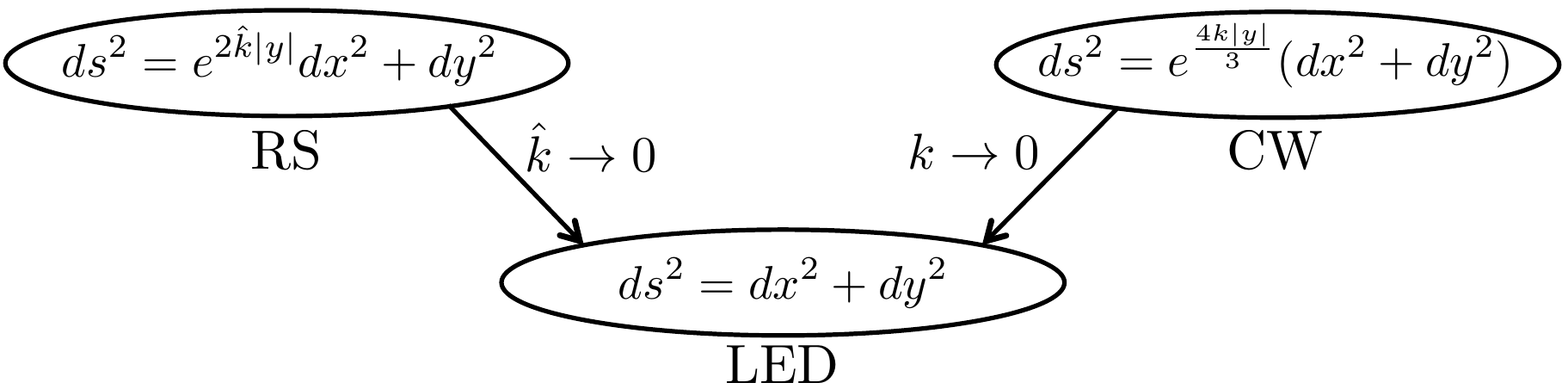}
\caption{A schematic view on how LED can be recovered as special limits of RS and the clockwork (CW). There is no simple one-parameter deformation to move between RW and CW.}
\label{fig:comp}
\end{figure}

As pictorially illustrated in  \fig{fig:comp}, both RS and the clockwork can be interpreted as deformations of LED, and the latter is recovered in the limit in which ${\hat k}$ or $k$ vanish. On the other hand, there is no simple one-parameter deformation from RS to the clockwork. This can also be understood at the level of fundamental theory, since turning the clockwork into RS cannot be done by modifying a parameter, but by suppressing a degree of freedom (the dilaton). Nonetheless, there are superficial, but important, similarities between the two theories, as evident for instance from the two corresponding expressions of $M_P$ in \eq{tabella}. However, these similarities hide a fundamental difference between the parameters $\hat k$ and $k$. In RS, $\hat k$ and $M_5$ are naturally of the same order. As shown in \Appf{frames}, both parameters are rescaled in the same way as we change coordinates from the $0$-frame to the $\pi$-frame, ${\hat k}'=e^{{\hat k} \pi R} {\hat k}$ and $M_5'=e^{{\hat k} \pi R} M_5$. Having $\hat k \ll M_5$ in RS is possible, but it looks like a tuning. On the other hand, in the clockwork, the parameter $k$ is protected by a dilaton shift symmetry. This difference shows up when we transform from the $0$-frame to the $\pi$-frame because, as shown in  \Appf{frames}, $k'=k$ and $M_5'=e^{2k \pi R/3} M_5$. From this perspective, it is evident that having $k \sim M_5$ in the clockwork is possible, but it looks like a coincidence. Since the two parameters are logically distinct, the natural expectation is that $k$ is smaller than $M_5$ (for $k> M_5$, the theory is not under control, as explained in \sect{sec:hierarchy}). Therefore, the onset of the gears could naturally appear at masses far below the cutoff $M_5$, unlike the case of RS.

The clockwork with $k<M_5$ reveals novel features, which are present in neither LED nor RS. For a fixed clockworking factor, the expression of the 4D Planck mass in \eq{tabella} is $M_P\sim M_5^{3/2}(2\pi R)^{1/2} e^{k\pi R}$. So the hierarchy is explained by a combination of volume and geometry, in an intermediate situation between LED and RS. The clockworking factor can then be smaller than its warping analogue, as it is assisted by a volume factor.
The KK interaction scale is also rather special in the clockwork. First of all, it is mode-dependent, decreasing with $n$ and saturating at large $n$. Second, it is larger than the 5D gravity scale by a factor $(M_5/k)^{1/2}$, which can be considerable for small $k$.

\sectioneq{Conclusions}
\label{sec:conclusions}
The clockwork is a highly efficient mechanism for generating exponentially suppressed interactions within a microscopic theory containing only $\mathcal{O}(1)$ parameters, in natural units, and a finite number of fields.  As a result, one can generate exponentially large interaction scales, even though no new physics appears at this high energy scale.  For this reason the clockwork has been used in models of inflation or relaxation to motivate apparent super-Planckian field excursions.  However, given that there are numerous clues for apparent new high interaction scales, such as the PQ scale, the neutrino see-saw scale, and even the Planck scale itself, it is interesting to investigate whether these interaction scales could in fact be a mirage, spawned by some form of clockwork mechanism.  The implication would be that all of the responsible new physics and the UV-completion are actually at much lower energies, possibly close to the TeV region.

\begin{figure}[t]
\centering
\includegraphics[height=2.5in]{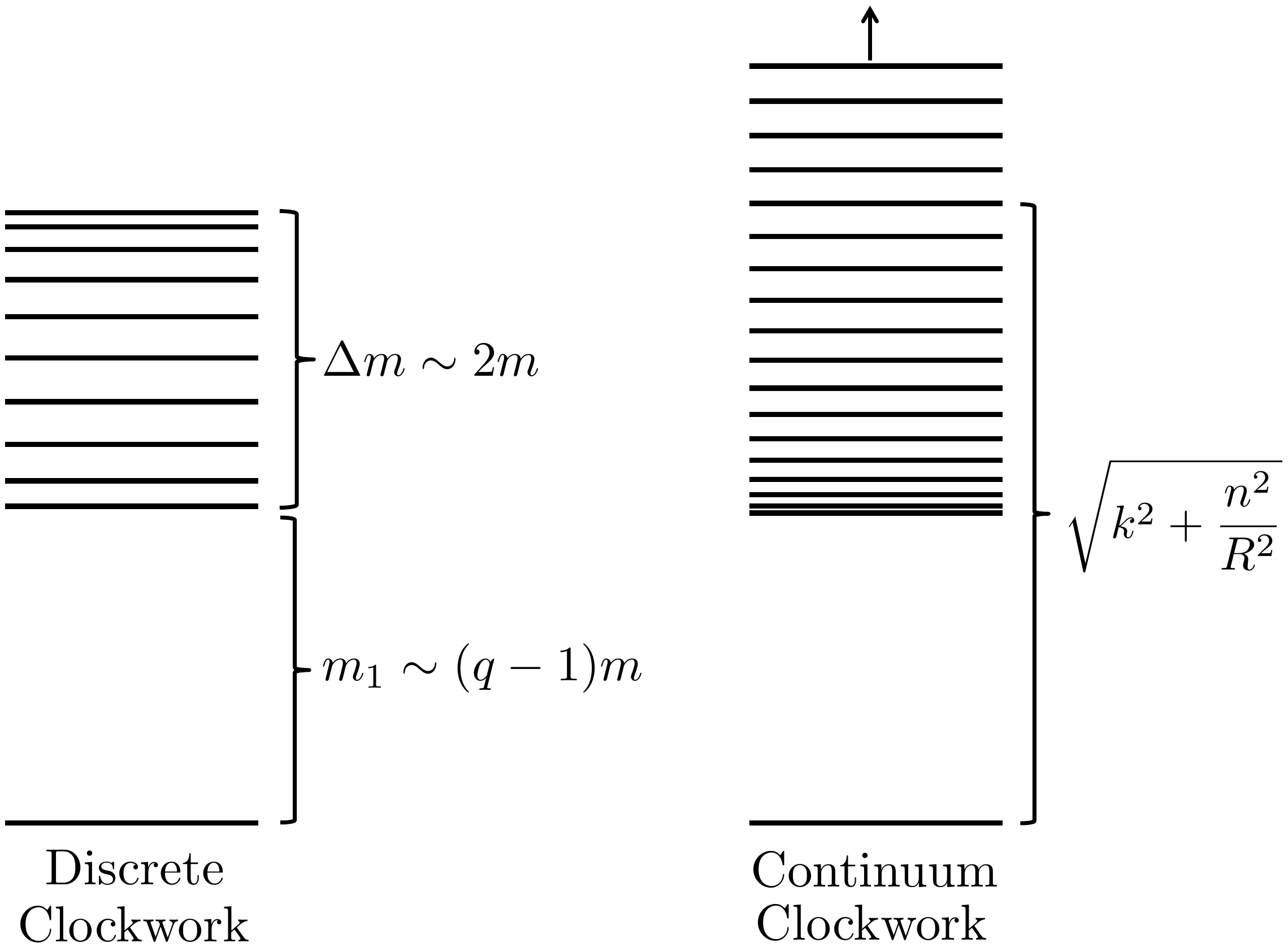}
\caption{The mass spectrum of the discrete (left) and continuum (right) clockwork.}
\label{fig:spec}
\end{figure}

In this work we have generalised the low energy structure of the scalar clockwork mechanism to fermions, vector bosons, and gravitons.  These models have obvious applications for the QCD axion, neutrino masses, flavour, dark sectors with millicharges, and multigravity theories.  In all these cases, exponentially large interaction scales can occur with all of the new physics at, or even below, the weak scale.  In every case the clockwork gives rise to a smoking gun spectrum of states, the clockwork `gears', shown in \fig{fig:spec}.  The peculiar spectrum of gears exhibits a mass gap, followed by a band of resonances whose couplings to the SM are \emph{not} suppressed, thus they could be discovered at colliders. 

As the clockwork mechanism can apply to a variety of fields, it is natural to search for an overarching clockwork framework.  To this end we studied the continuum limit $N \to \infty$ in which the clockworking factor $q^N$ remains exponentially large, but finite.  A setup with site-independent mass parameters was studied. We expect that the results would not change qualitatively if different masses of comparable size were considered, although it may be interesting to generalise our findings to alternative lattices and corresponding geometries.

 We found that the continuum clockwork corresponds to a 5D theory which has previously been studied in the context of linear dilaton duals to Little String Theory \cite{Lykken:1996fj,Antoniadis:1999rm,Antoniadis:2001sw,Cox:2012ee}.  This reveals a resolution of the hierarchy problem which has some features in common with LED and RS, but is conceptually distinct from these extra dimensional theories.  In particular, as shown in \fig{fig:spec}, the spectrum of resonances differs from either LED or RS models, with a mass gap controlled by one parameter, followed by a series of excitations controlled by a separate parameter.  This leads to a distinctive collider phenomenology where, depending on experimental resolution, the band of new resonances may show up as individual particles or as a fat continuum contribution, as was considered in \cite{Baryakhtar:2012wj} for the linear dilaton theory. The collider phenomenology of both the discrete and continuum versions of the clockwork will be presented in forthcoming work.
 
\bigskip
\subsubsection*{Acknowledgements}

We are very grateful to Riccardo Rattazzi for many enlightening discussions and for his precious and generous suggestions. We wish to thank Asimina Arvanitaki, Masha Baryakhtar, Savas Dimopoulos, Tony Gherghetta, Anson Hook, John March-Russell, Marek Olechowski, Stefan Pokorski, Antonio Riotto, Prashant Saraswat, Julian Sonner, Matt Strassler, and Alfredo Urbano for useful discussions and comments.

\appendix

\sectioneq{Useful formul{\ae} for 5D gravity}
\label{sec:5D}

For the ease of the reader we collect some general formul{\ae} that are useful for projecting 5D gravity into 4D.
Let us consider a general metric of the form
\be
ds^2 = g_{MN}\, dx^M dx^N= \hat{g}_{\mu\nu} (x,z) \, dx^\mu dx^\nu + dz^2 ~~.
\ee
The 5D gravitational action is
\bea
&&{\mathcal S} =  \int d^5x  \sqrt{-g} \, \frac{M_5^3}{2} \,  \mathcal{R}_5 (g) =
\label{eq:fullaction} \\
&&\! \! \!\! \! \int d^4x \, dz \sqrt{-\hat{g}} \, \frac{M_5^3}{2} \left[ \mathcal{R}_4 (\hat{g}) +\frac{ (\hat{g}^{\mu\nu} \partial_z \hat{g}_{\mu\nu})^2}{4} +\frac{(\partial_z \hat{g}^{\mu\nu}) (\partial_z \hat{g}_{\mu\nu})}{4}  
- \frac{1}{\sqrt{-\hat{g}}} \partial_z \left( \sqrt{-\hat{g}} \hat{g}^{\mu\nu} \partial_z \hat{g}_{\mu\nu} \right) \right] \, ,
\nonumber
\eea
where $\mathcal{R}_5 (g)$ and $\mathcal{R}_4 (\hat{g})$ are the 5D and 4D Ricci scalar. The last term in \eq{eq:fullaction} is a total derivative, thus to determine the graviton properties we may discard it.  However, we have included it above as it enables a quick determination of the Ricci scalar, since the term in the square brackets is the effective 5D Ricci scalar.  

Let us now focus on metrics defined by
\beq
ds^2 = e^{2{\hat \sigma}(z)}\tilde{g}_{\mu\nu} (x)\, dx^\mu dx^\nu + dz^2=
e^{2{ \sigma}(y)}\left[ \tilde{g}_{\mu\nu} (x) \, dx^\mu dx^\nu + dy^2\right]
\label{supermetric}
\eeq
The second form is obtained after a change of variables such that $dz/dy = e^{ \sigma}$ and $\sigma(y) = \hat{\sigma} (z)$ when $z$ is expressed in terms of $y$ or vice versa.  This implies
\beq
\hat \sigma' =\sigma' \, e^{-\sigma} ~,~~~~~~\hat \sigma'' =(\sigma''-\sigma'^2) \, e^{-2\sigma} ~,
\eeq
where the primes denote derivatives with respect to the corresponding variable ($z$ for $\hat \sigma$ and $y$ for $\sigma$).

From now on, in the rest of this appendix, we will set the intrinsic 4D metric to be flat, taking $\tilde{g}_{\mu\nu}=\eta_{\mu \nu}$. Then,
the 5D Ricci scalar becomes
\beq
\mathcal{R}_5 =-4(2{\hat \sigma}'' +5 {\hat \sigma}'^2) =-4(2{\sigma}'' +3 { \sigma}'^2) e^{-2 \sigma}~~.
\eeq

We are especially interested in the cases of the clockwork (CK) and warped (RS) spaces, in which the functions $\sigma (y)$ and $\hat \sigma (z)$, defined in the ``upper part" of the orbifold with $y_0 \le y \le y_\pi$ and $z_0 \le z \le z_\pi$ (with the ``lower part" obtained by orbifold symmetry), are given by
\beq
 \sigma_{CW}(y) = \frac 23 k y ~,~~~y_0=0~,~y_\pi =\pi R~,~~~~
{\hat \sigma}_{CW}(z) = \ln \frac{z}{z_0} ~,~~~z_0=\frac{3}{2k}~,~z_\pi =z_0e^{\frac 23 k\pi R}~,
\label{metCW}
\eeq
\beq
{\hat \sigma}_{RS}(z) = {\hat k}z ~,~~~z_0=0~,~z_\pi =\pi R~,~~~~
 \sigma_{RS}(y) = -\ln (1-{\hat k}y) ~,~~~y_0=0~,~y_\pi =\frac{1-e^{{\hat k}\pi R}}{\hat k}~.
 \label{metRS}
\eeq
 
Thus, we find the curvature of the clockwork and AdS$_5$ spaces 
\bea
{\rm CW}  ~~&\Rightarrow & ~~~~~ \mathcal{R}_{CW}  =   - \frac{16}{3}\, k^2 e^{-4 k y/3}=- \frac{16}{3}\left( \frac{kz_0}{z}\right)^2 ~,
\label{curvatureCW} \\
{\rm RS}  ~~&\Rightarrow & ~~~~~\mathcal{R}_{RS}   =   - 20 \, {\hat k}^2 ~.
\label{curvatureRS}
\eea
These expressions show that the clockwork space has the geometry of a cone with a singularity at $z=0$ ($y=-\infty$). The compactification of the extra dimension selects a slice of the conical space in which the singularity is avoided.\footnote{Such conical singularities may arise in string compactifications (see e.g. \cite{Klebanov:1998hh,Klebanov:2000hb}).}  For RS, we recognise the AdS$_5$ space with constant curvature.

The Einstein equation is $\mathcal{G}_{MN} =T_{MN}/M_5^3$, where the Einstein tensor in the two coordinate choices is given by
\bea
\mathcal{G}_{MN} & = & \mathcal{R}_{MN} - \frac{1}{2} g_{MN} \mathcal{R}  
\nonumber \\ & = &
3({\hat \sigma}''+2{\hat \sigma}'^2)e^{2{\hat \sigma}} (\eta_{MN}-\delta_{M5}\delta_{N5} ) + 6{\hat \sigma}'^2 \delta_{M5}\delta_{N5} 
\nonumber \\ & = &
3({ \sigma''}+{ \sigma}'^2) (\eta_{MN}-\delta_{M5}\delta_{N5} ) + 6{ \sigma}'^2 \delta_{M5}\delta_{N5} ~.
\label{Einstein}
\eea
The energy-momentum tensor for a canonically-normalised real scalar field $\Phi$ is
\bea
T_{MN} &=&\partial_M \Phi \, \partial_N \Phi -g_{MN} \left( \frac 12\, g^{PQ}\, \partial_P \Phi \, \partial_Q \Phi +V\right)
\nonumber \\
&=& -\left[ \frac{(\partial_z \Phi)^2}{2}+V \right] e^{2{\hat \sigma}} (\eta_{MN}-\delta_{M5}\delta_{N5} ) + \left[ \frac{(\partial_z \Phi)^2}{2}-V \right] \delta_{M5}\delta_{N5} \nonumber \\
&=& -\left[ \frac{(\partial_y \Phi)^2}{2}+e^{2 \sigma} V \right]  (\eta_{MN}-\delta_{M5}\delta_{N5} ) + \left[ \frac{(\partial_y \Phi)^2}{2}-e^{2 \sigma} V \right] \delta_{M5}\delta_{N5}
~,
\eea
where $V(\Phi)$ is the scalar potential for the field $\Phi$. In the last two lines we have written the energy momentum tensor for the metrics in \eq{supermetric}, under the assumption that the background value of $\Phi$ is independent of the 4D coordinates.

The equation of motion of the scalar $\Phi$ is
\beq
\frac{1}{\sqrt{-g}}\partial_M \sqrt{-g}\, g^{MN}\partial_N \Phi =\frac{dV}{d\Phi} ~,
\eeq
which, for the two choices of coordinates, becomes
\beq
\left( \partial_z^2 +4 {\hat \sigma}' \partial_z \right) \Phi = \frac{dV}{d\Phi} ~,
\eeq
\beq
\left( \partial_y^2 +3 { \sigma}' \partial_y \right) \Phi = e^{2 \sigma} \frac{dV}{d\Phi} ~.
\eeq

We can also derive the relationship between the 5D and 4D Planck masses. From \eq{eq:fullaction} we extract the value of $M_P$ as the coefficient in front of the 4D Einstein-Hilbert action
\beq
M_P^2 =2M_5^3 \int_{z_0}^{z_\pi} dz\, e^{2\hat \sigma} =  2M_5^3 \int_{y_0}^{y_\pi} dy\, e^{3 \sigma}~,
\eeq
where $z_{0, \pi}$ and $y_{0, \pi}$ are the coordinates of the orbifold fixed points in the two frames. For RS and the clockwork, using eqs.~(\ref{metCW})--(\ref{metRS}) we find
\bea
{\rm CW}  ~~&\Rightarrow & ~~~~~
M_P^2= \frac{M_5^3}{k} \left( e^{2k\pi R} -1 \right)~,
\\
{\rm RS} ~~&\Rightarrow &~~~~~
M_P^2=\frac{M_5^3}{{\hat k}} \left(e^{2{\hat k}\pi R} -1\right) ~.
\eea
The two expressions are identical, once we identify $k$ with $\hat k$.

\sectioneq{Graviton action}
\label{sec:gravaction}
Let us start with the 5D gravity action
\beq
{\mathcal S} =  \int d^5x  \sqrt{-g} \left( \frac{M_5^3}{2} \,  \mathcal{R}_5 + \mathcal{L}_M \right) ~,
\eeq
where $\mathcal{L}_M$ is the matter Lagrangian in the bulk.
We will consider the $y$ coordinates defined in the second metric of \eq{supermetric} where $\tilde{g}_{\mu\nu}$ can be thought of as containing fluctuations about 4D Minkowski space.  We discard the total derivative term in \eq{eq:fullaction} and expand the metric as 
\be
\tilde{g}_{\mu\nu} =  \eta_{\mu\nu}+\frac{2}{M_5^{3/2}} \, h_{\mu\nu}  ~~~~,~~~~ \tilde{g}^{\mu\nu} =  \eta^{\mu\nu}-\frac{2}{M_5^{3/2}}\,  h^{\mu\nu}+\frac{4}{M_5^{3}} \, h^{\mu\lambda} h_{\lambda}^{\nu} +{\mathcal O}(h^3) ~~.
\ee
By choosing the `transverse-traceless' gauge $\partial_\mu h^{\mu\nu} = 0$, $\eta_{\mu\nu} h^{\mu\nu} = 0$, to lowest order in $h$, the action becomes
\bea
S  & = & \int d^4x \, dy\,  e^{3{ \sigma}} \bigg[-\frac{1}{2} ( \partial_{\lambda} h_{\mu\nu}) ( \partial^{\lambda} h^{\mu\nu}) -\frac{1}{2} (\partial_y h_{\mu\nu}) (\partial_y h^{\mu\nu}) \\
& &- 6  \sigma' h_{\mu\nu} \partial_y h^{\mu\nu}- \left( 6 {\sigma}'^2 +\frac{e^{2{ \sigma}}}{M_5^{3}} \mathcal{L}_M \right) h_{\mu\nu}  h^{\mu\nu} \bigg] ~~, \nonumber
\eea
where the last term arises from the expansion of $\sqrt{-{g}}$.  Through integration by parts the last two terms may be rearranged to give 
\bea
S  & =  & \int d^4x \, dy\, e^{3{ \sigma}} \bigg[-\frac{1}{2} ( \partial_{\lambda} h_{\mu\nu}) ( \partial^{\lambda} h^{\mu\nu})-\frac{1}{2} (\partial_y h_{\mu\nu}) (\partial_y h^{\mu\nu}) 
\label{eq:almostfinalaction}
\\
& &  + \left( 3 (\sigma'' +{ \sigma}'^2 ) -\frac{e^{2{ \sigma}}}{M_5^{3}}  \mathcal{L}_M \right)  h_{\mu\nu}  h^{\mu\nu} \bigg] ~~.
\nonumber
\eea
If we assume that the bulk matter configuration does not depend on 4D coordinates, than the energy momentum-tensor that sources the metric is $T_{\mu\nu} = g_{\mu\nu} \mathcal{L}_M$. Using \eq{Einstein}, the 4D components of the Einstein equation $\mathcal{G}_{\mu \nu} =T_{\mu \nu}/M_5^3$ give
\beq
3(\sigma '' +\sigma'^2 ) = \frac{e^{2{ \sigma}}}{M_5^{3}} \mathcal{L}_M ~.
\eeq
Thus, on the equations of motion, the second line in \eq{eq:almostfinalaction} vanishes and we are left only with the first line, which is the result used in \eq{eq:finalaction}.

\sectioneq{Two equivalent frames for the clockwork}
\label{sec:frames}

The discrete clockwork Lagrangian, in all its versions (scalar, fermion, vector, and graviton), has a discrete parametrisation invariance under which the parameters transform as $q\to q' =1/q$, $m \to m' = qm$, and the site $j$ is exchanged with the site $j'=N-j$. This means that any clockwork theory with a given $q>1$ is physically equivalent to a theory with $0<q<1$, once the mass parameter is appropriately transformed and the role of the two end sites is reversed. We will call ``$N$-frame" the representation with $q>1$ and ``$0$-frame" the representation of the same theory with $q<1$. 

In our study, we have adopted the $N$-frame, in which the zero-mode component is exponentially suppressed at the site $N$, as shown by \eq{rotation}
\beq
O_{j0} = \frac{1}{q^j} \left[ 1+ {\mathcal O}\left( \frac{1}{q^2}\right) \right] ~~\Rightarrow ~~ O_{00}\approx 1
~,~~O_{N0}\approx q^{-N} \ll 1 ~~~(N\mbox{-frame}~,~~q>1)~.
\eeq
In the equivalent ``$0$-frame", the parameter $q$ is smaller than one and the zero mode is exponentially suppressed at the site $0$. Indeed, from \eq{rotation} in the limit of small $q$, we find
\beq
O_{j0} = q^{N-j} \left[ 1+ {\mathcal O}\left( q^2\right) \right] ~~\Rightarrow ~~ O_{00}\approx q^N \ll 1
~,~~O_{N0}\approx  1 ~~~(0\mbox{-frame}~,~~q<1)~.
\eeq
While in the $N$-frame the clockwork mechanism operates when the external sector is coupled at the site $N$, in the $0$-frame the coupling must occur at the site $0$. However, 
the two frames are physically equivalent and the choice is purely a matter of convention or of phenomenological convenience.

The same discrete parametrisation invariance found in the discrete clockwork also appears in the continuum clockwork. For convenience, let us consider only the ``upper part" of the orbifold parametrised by $0<y < \pi R$, while the ``lower part" ($- \pi R<y<0$) can be easily recovered by orbifold symmetry ({\it i.e.} by requiring invariance under $y\to -y$). The parametrisation invariance in the continuum is $k\to k'=-k$ and $y\to y' =\pi R -y$. We call ``$0$-frame" the case  in which the metric in \eq{cmetric} has $k>0$ and ``$\pi$-frame" the case with $k<0$. In this paper we have adopted the $0$-frame in which, according to \eq{eigen1}, the zero-mode wavefunction and its probability density are
\beq
\psi_0 (y) =  \sqrt{\frac{k \pi R}{ e^{2 k \pi R} -1 }} ~~\Rightarrow ~~\frac{dP}{dy} \approx k\, e^{2k(y-\pi R)} ~~~(0\mbox{-frame}~,~~k>0)~.
\eeq
In the ``$\pi$-frame" we find
\beq
\psi_0 (y) =  \sqrt{\frac{|k| \pi R}{ 1- e^{-2 |k| \pi R} }} ~~\Rightarrow ~~\frac{dP}{dy} \approx |k|\, e^{-2|k|y} ~~~(\pi\mbox{-frame}~,~~k<0)~.
\eeq
In the two (physically equivalent) frames, the role of the two branes is switched. In the $0$-frame, the clockwork mechanism operates when we couple the external sector at $y=0$, where the zero mode is exponentially suppressed. In the $\pi$-frame, the external sector must be localised on the brane at $y=\pi R$. The exponential suppression of the effective coupling is the same in the two frames.

Let us now consider the full 5D construction with the metric consistently induced by the dynamics. For an instructive comparison, let us first consider the case of RS which, in the $0$-frame, is described by the action
\bea
S({{\hat k}},R,M_5,m,y_{0,\pi})&=&2\int_0^{\pi R} dy\, \int d^4 x \sqrt{-g} \left[ \frac{M_5^3}{2}( {\mathcal R} +12 {\hat k}^2 ) 
\right. \nonumber \\ 
 &+& \left. \frac{\delta (y-y_0)}{\sqrt{g_{55}}} \left( {\mathcal L}(\Phi , m)+6M_5^3{{\hat k}}  \right)  -\frac{\delta (y-y_\pi)}{\sqrt{g_{55}}}  6M_5^3{{\hat k}}  \right] ~.
\eea
Here $y_0=0$ and $y_\pi =\pi R$ are the brane locations, ${\mathcal L}$ is the localised matter Lagrangian involving a set of fields $\Phi$ and mass parameters $m$.

To obtain the equivalent action in the $\pi$-frame, we perform the change of coordinates 
\beq
\frac{y}{R}= \pi - \frac{y'}{R'} ~,~~~~~~{\rm with}~\frac{R}{R'}=w~,~~~~w=e^{{{\hat k}} \pi R}~,
\eeq
under which the line elements becomes
\beq
ds^2 = e^{2{{\hat k}} y} \, dx^2 + dy^2=w^2\left( e^{-2{\hat k}' y'} \, dx^2 + dy'^2 \right)~,~~~~~~{\hat k}' = w\, {{\hat k}}~.
\eeq
Note that ${\hat k}'$ has been defined such that ${\hat k}'R' = {{\hat k}}R$, and therefore the warping factor $w$ is the same in both frames. This change of coordinates implies the transformations
\bea
g_{MN} \to w^2 \, g'_{MN}~,~~
\sqrt{-g}\, dy \! \! \! &\to& \! \! \! w^5\, \sqrt{-g'} \, dy' ~,~~
{\mathcal R} (g) \to w^{-2}\,  {\mathcal R}' (g') ~,
\nonumber \\
\frac{\delta (y-y_{0,\pi})}{\sqrt{g_{55}}}\, dy\! \! \! &\to& \! \! \! w^{-1} \,  \frac{\delta (y'-y'_{\pi ,0})}{\sqrt{g'_{55}}}\, dy' ~,
\eea
where $g_{MN}$ is the usual RS metric in the $0$-frame, $g'_{MN}$ is its inverse involving primed quantities ({\it i.e.} $g'$ is obtained from $g$ with the replacement ${{\hat k}}y \to -{\hat k}'y'$), and  $y'_0=0$, $y'_\pi =\pi R'$. 

We can compensate the rescaling of the metric by defining the rescaled fields $\Phi'$ (where $\Phi =\phi$, $\psi$, $A_\mu$ for scalar, fermion, or gauge fields) and rescaled mass parameters $m'$
\beq
\phi' =w\, \phi  ~,~~
\psi' = w^{3/2}\, \psi ~,~~
A'_\mu = A_\mu ~,~~
m' = w \, m ~,~~
\eeq
such that
\beq
{\mathcal L}(\Phi , m)=w^{-4}\, {\mathcal L}(\Phi' , m') ~.
\eeq
Finally, with the definition
\beq
M'_5=w \, M_5 
\eeq
we obtain that the relation between the actions in the $0$ and $\pi$-frames is
\beq
S({{\hat k}},R,M_5,m,y_{0,\pi})=S(-{\hat k}',R',M'_5,m',y'_{\pi ,0}) ~.
\eeq

This shows the equivalence of the $0$-frame action and the $\pi$-frame action obtained by multiplying all mass parameters  by a warping factor $w$, inverting the metric (${{\hat k}}\to -{\hat k}'$), and reversing the role of the two branes ($y_{0,\pi}\to y'_{\pi ,0}$). In the $0$-frame, the SM is coupled at $y_0$, the mass parameters are of the order of the weak scale and the 4D Planck mass is $M_P^2 \approx (M_5^3/{{\hat k}}) w^2$. In the $\pi$-frame, the SM is coupled at $y'_\pi$, the mass parameters are of the order of the UV scale and the 4D Planck mass is $M_P^2 \approx {M'}_5^3/{\hat k}'$. In either case, the warping factor $w$ is given by the ratio of the Planck to weak scale.

Let us now consider the case of the clockwork, whose action in the $0$-frame is
\bea
S( k,R,M_5,m,y_{0,\pi})&=&2\int_0^{\pi R} dy\, \int d^4 x \sqrt{-g} \left[ \frac{M_5^3}{2}\left( {\mathcal R} -\frac{1}{3} \, g^{MN} \partial_M S\, \partial_N S + e^{-\frac{2S}{3}}\, 4k^2 \right) 
\right. \nonumber \\ 
 &+& \left. 
 \frac{\delta (y-y_0)}{\sqrt{g_{55}}} \left( {\mathcal L}(\Phi , m) + 4e^{-\frac{S}{3}}M_5^3 k  \right)  -\frac{\delta (y-y_\pi)}{\sqrt{g_{55}}}  4e^{-\frac{S}{3}}M_5^3 k  \right] ~.
\eea
Note that we have introduced the SM Lagrangian ${\mathcal L}$ in the Einstein frame with no direct coupling to $S$. In the $0$-frame, different choices of coupling ${\mathcal L}$ in the Einstein or Jordan frames, with or without $S$ couplings, are all equivalent since $S(y_0)=0$. However, the choice matters when we compare the effects on the two branes. Our choice ensures that SM couplings do not depend on the dilaton background. 
 
We can now obtain the equivalent $\pi$-frame description with the change of coordinates
\beq
y=\pi R - y'
\eeq
such that the line element becomes
\beq
ds^2 = e^{\frac 43 ky} (dx^2+dy^2) =w^{\frac 43}\, e^{-\frac 43 ky'} (dx^2+d{y'}^2)
~,~~~~w=e^{k \pi R}~.
\eeq
This implies the transformation
\beq
g_{MN} \to w^{\frac 43} \, g'_{MN}~,
\eeq
where, as before, $g'$ is the inverse of $g$ in the new coordinates ({\it i.e.} obtained with the replacement $ky \to -ky'$). With the redefinition
\beq
S' = S-\ln w^2 ~,~~~~M'_5= w^{\frac 23} \, M_5
\eeq
and the field and mass rescaling
\beq
\phi' =w^{\frac 23}\, \phi  ~,~~
\psi' = w\, \psi ~,~~
A'_\mu = A_\mu ~,~~
m' = w^{\frac 23} \, m ~,~~
\eeq
such that
\beq
{\mathcal L}(\Phi , m)=w^{-\frac 83}\, {\mathcal L}(\Phi' , m') ~,
\eeq
we find that the transformed action is related to the original one by
\beq
S( k,R,M_5,m,y_{0,\pi})=S(-k,R,M'_5,m',y_{\pi ,0}) ~.
\eeq
This equation shows the equivalence of the actions in the $0$ and $\pi$-frames.

For the clockwork, the $\pi$-frame is obtained by inverting the metric ($k \to -k$), reversing the role of the branes ($y_{0,\pi}\to y_{\pi ,0}$), rescaling the 5D Planck mass $M_5$ and the Higgs mass $m$ by a factor $w^{\frac 23}$, {\it but leaving $|k|$ and $R$ invariant}. This last feature is an important difference with respect to RS. In RS, $M_5$, $m$, ${\hat k}$, and $1/R$ are all rescaled equally as we change frame, so they are all expected to be of the order of the cutoff scale. On the other hand, in the clockwork, the parameters $k$ and $R$ do not rescale and thus
 the masses of the clockwork gears are typically not correlated with the cutoff scale $M_5$. This is because $k$ is protected by a shift symmetry of $S$ in the bulk and can be naturally smaller than $M_5$.
 
The 4D Planck mass is given by $M_P^2 \approx {M'}_5^3/k=({M}_5^3/k)w^2$, so the clockworking factor $w$ corresponds to the ratio between Planck and weak masses, just as in RS. However, unlike RS, the 5D Planck mass is rescaled only by the factor $w^{\frac 23}$.

\sectioneq{Goldberger-Wise radius stabilisation}
\label{sec:GW}

As discussed in \Sec{secgeometry}, an attractive feature of clockwork gravity induced by a dilaton in a 5D space with an extra dimension compactified on a $S_1/Z_2$ orbifold is that the radius $R$ of the extra dimension can be naturally stabilised  at values $kR ={\cal O} (1)$ with boundary conditions for the dilaton on the branes alone. We want to show here that the stabilisation mechanism proposed by Goldberger and Wise~\cite{Goldberger:1999uk} for warped geometry can also work for the clockwork geometry, although it is less economical in terms of field content than dilaton boundary conditions.

In order to generate a non-trivial potential for the radion mode $R$, let us introduce
 a real 5D scalar field $\varphi$ with mass $m_\varphi$ and add to the Jordan-frame action in \eq{actionstring} the term
\beq
{\mathcal S}=-\frac 12 \int d^5x \, \sqrt{-g}\, e^S \left( g^{MN}\, \partial_M \varphi \, \partial_N \varphi + m_\varphi^2\, \varphi^2 \right) ~.
\label{actiongw}
\eeq
We are assuming that the mass $m_\varphi$ is sufficiently small, so that the field $\varphi$ does not modify the underlying metric nor the dilaton profile. On this static background, and after performing the transformation in \eq{tram} to go to the Einstein frame, the action for $\varphi$ becomes
\beq
{\mathcal S}=-\frac 12 \int d^4x\, dy \, e^{2k|y|}\left[ (\partial_\mu \varphi )^2 +  (\partial_y \varphi )^2 + m_\varphi^2\, \varphi^2 \right] ~.
\label{actiongw2}
\eeq

For configurations that do not depend on 4D space-time coordinates, the equation of motion of $\varphi$ is
\beq
\left( e^{-2k|y|}\, \partial_y \, e^{2k|y|}\,  \partial_y - m_\phi^2 \right) \varphi =0 ~.
\eeq
The most general solution is
\beq
\varphi (y)= A_+\,  e^{(\nu -1)k|y|}+A_- \, e^{-(\nu +1)k|y|} ~,~~~~~~\nu \equiv \sqrt{1+\epsilon} ~, ~~~~~~\epsilon \equiv \frac{m_\varphi^2}{k^2} ~.
\eeq
The integration constants $A_\pm$ are fixed by the Dirichlet boundary conditions on the branes $\varphi (0) =\varphi_0$ and $\varphi (\pi R) =\varphi_\pi$, which give
\beq
A_\pm = \frac{ \varphi_\pi e^{(1\pm \nu ) k \pi R}-\varphi_0}{e^{\pm 2 \nu k \pi R} -1} ~.
\eeq
In a complete dynamical model, the values of $\varphi_{0,\pi}$ are expected to come from interactions localised on the branes, but their origin is not essential for our discussion.

The radion potential $V(R)$ is obtained by integrating \eq{actiongw2} over the extra dimension
\bea
V(R) &=&  \int_0^{\pi R} dy \, e^{2ky} \left[ (\partial_y \varphi )^2 + m_\varphi^2\, \varphi^2 \right]
\nonumber \\
&=& k\left[ A_+^2 (\nu -1 )  \left( e^{ 2\nu k\pi R}-1 \right) +A_-^2 (\nu +1)  \left( 1- e^{ -2\nu k\pi R} \right) \right] ~.
\eea
Since we are interested in the limit of small $m_\varphi$, we can expand the result in powers of $\epsilon$, taking however $\epsilon k\pi R \sim {\mathcal O}(1)$. At leading order, we obtain
\beq
V(R) = 2k \left( \varphi_\pi e^{-\frac{\epsilon k\pi R}{2}} -\varphi_0\right)^2 ~.
\eeq
Since this potential is never negative, its minimum is reached when the term in brackets vanishes. This corresponds to the value at which the compactification radius is stabilised,
\beq
kR= \frac{2}{\pi \epsilon} \ln \frac{\varphi_\pi}{\varphi_0} ~.
\label{stabile}
\eeq
For parameters of order unity and a moderately small $\epsilon$, we naturally obtain values of $kR$ that can explain the hierarchy between the weak and gravity scales. 
For instance, for $\varphi_\pi / \varphi_0 =e$ and $\epsilon =0.06$, we find $kR =10$, in agreement with \eq{eccokR}.

\sectioneq{Deconstructing the clockwork dimension}
In our study we started from the discrete clockwork and used the $N\! \to \! \infty$ limit to motivate the metric of the continuum clockwork. Here we want to conclude our itinerary by taking the reverse path to show how the deconstruction of the clockwork dimension leads to the same low energy theory as the discrete models discussed in \sect{secmechanism}.

The cases of the scalar and vector clockworks are relatively simple, thus we will treat them together. The action of a scalar and vector field in the 5D clockwork space is
\be
{\mathcal S} = -\frac 12 \int d^4x \,  \int_{\-\pi R}^{\pi R} dy  \left[  e^{2 k |y|} \, (\partial_M \phi)^2 + \frac{1}{2} \, e^{\frac 23 k |y|}\, F_{MN}^2  \right] ~~.
\ee
For convenience, in this appendix we will let the fifth coordinate vary in the interval $0<y<\pi R$ and absorb the extra factor of 2 in the action with a field redefinition. 
After decomposing the 5D indices and with a convenient field rescaling, we find
\be
{\mathcal S} = -\frac 12 \int d^4x \int_0^{\pi R} dy \left[ (\partial_\mu \phi)^2   + \frac{1}{2} F_{\mu\nu}^2  +   e^{2 k y}  (\partial_y\, e^{-k y} \phi)^2 + e^{\frac 23 k y} (\partial_y \,  e^{-\frac 13 k y} A_\mu)^2 \right]  ~.
\ee
Here we have also assumed Dirichlet boundary conditions for the 5D component of the gauge field, such that it does not propagate.  Now let us discretise the fifth dimension, such that $y = j a$ and $\pi R = Na$,
\bea
{\mathcal S}  & = & - \frac 12 \int d^4x \bigg\{ \sum_{j=0}^{N} \left[  (\partial_\mu \phi_j)^2   + \frac{1}{2} F_{j\, \mu\nu}^2   \right] + \\
& & 
\frac{1}{a^2} \sum_{j=0}^{N-1} \left[  \left( \phi_j - e^{-k a} \phi_{j+1} \right)^2 + \left( A_{\mu\, j} -  e^{-\frac 13 k a} A_{\mu\, j+1}  \right)^2   \right] \bigg\} ~~. \nonumber
\eea
Flipping the sign of $k$ to go to the $\pi$-frame, we conclude that
the deconstruction realises the scalar and vector discrete clockwork models with 
\beq
m_\phi = \frac{N}{\pi R} ~,~~~q_\phi = e^{\frac{k\pi R}{N}} ~~{\rm and}~~~~~m_A =\frac{N}{\pi R}~,~~~q_A = e^{\frac{k\pi R}{3N}}  ~.
\eeq

Let us now consider fermions.  The action of a massless fermion in the 5D clockwork space-time is
\be
{\mathcal S} = -\int d^4 x\,  \int_{\-\pi R}^{\pi R} dy\, e^{\frac 83 k y}\, \frac{i}{2} \left( \overline{\psi} \gamma^M \overset\leftrightarrow{\partial}_M \psi \right) ~~,
\ee
where $\psi$ is a 4-component spinor in 5D, $\gamma^M =(\gamma^\mu, i\gamma_5)$, and $\overset\leftrightarrow{\partial} = \overset\rightarrow{\partial}-\overset\leftarrow{\partial}$ with derivatives acting only inside the parenthesis. The spin connection can be dropped from the action as, although it is non-zero, its contributions for a metric of this form cancel (see e.g. \cite{Gherghetta:2000qt,Csaki:2003sh,Csaki:2007ns} for related discussions). Decomposing the 5D indices and projecting the spinor in its left and right components using the 4D chiral projector, we obtain

\be
{\mathcal S} = \int d^4 x \,  \int_{\-\pi R}^{\pi R} dy\,  e^{\frac 83 k y} \left[ -i \overline{\psi} \gamma^\mu \partial_\mu \psi
+\frac 12 \left(  \overline{\psi}_L \, \partial_y \psi_R -\partial_y \overline{\psi}_L \, \psi_R +\hc \right) \right] ~~.
\ee

When discretised, this action leads to a doubling of the zero modes, in a similar manner as the fermion doubling problem in lattice gauge theories.  Following \cite{Bai:2009ij} we cure this problem by adding a Wilson term
\beq
{\mathcal S}_W = -  
\int d^4 x\,  \int_{\-\pi R}^{\pi R} dy \, e^{\frac 83 k y}\, \frac{\eta a}{2} \, \partial_y \overline{\psi} \, \partial_y \psi =
-\int d^4 x \,  \int_{\-\pi R}^{\pi R} dy \, e^{\frac 83 k y}\,  \frac{\eta a}{2}\left( \partial_y \overline{\psi}_L \, \partial_y \psi_R  + \hc \right)
~~.
\eeq
This operator is higher dimensional, and thus vanishes in the continuum.  It is introduced in order to remove one of the hopping directions, which would otherwise give rise to the usual fermion doubling problem.  

We can now rescale the field $\psi \to e^{-\frac 43 k y} \psi$ and
 discretise the theory to obtain
\bea
&&{\mathcal S} + {\mathcal S}_W  =  \int d^4 x \sum_{j=0}^N - i  \left(  \overline{\psi}_{Lj} \gamma^\mu \partial_\mu \psi_{Lj} 
+  \overline{\psi}_{Rj} \gamma^\mu \partial_\mu \psi_{Rj}\right) 
 \\
& +&  \frac{1}{2a} \sum_{j=0}^{N-1} 
 \left[ (1+\eta)e^{-\frac 43 k a} \overline{\psi}_{Lj} \psi_{Rj+1}- (1-\eta )e^{-\frac 43 k a}  \overline{\psi}_{Lj+1} \psi_{Rj} -\eta (1+e^{-\frac 83 ka}) \overline{\psi}_{Lj} \psi_{Rj}+ \hc \right] \nonumber
\eea
Here we have eliminated some $\eta$-dependent terms of the form $\overline{\psi}_N \psi_N$ and $\overline{\psi}_0 \psi_0$ with appropriate counterterms localised at the boundaries. 

As desired, with the particular choice of $\eta= \pm 1$ the Wilson operator allows for one of the hopping directions to be removed, addressing the fermion doubling problem. Taking $\eta =1$ we recover the discrete clockwork Lagrangian in \eq{lagfer} in the $\pi$-frame with 
\beq
m_\psi =\frac{N}{\pi R} ~,~~~q_\psi = e^{\frac{4k\pi R}{3N}} ~,
\eeq
up to corrections subleading in $1/N$.

Finally, let us consider the clockwork gravitons. (For deconstructions of gravity in general scenarios see \cite{ArkaniHamed:2002sp,Schwartz:2003vj} and for RS see \cite{Randall:2005me}.)
Discretisation of the fifth dimension in the action in \eq{eq:finalaction} leads to a graviton action
\bea
{\mathcal S}  & = & -\frac 12  \int d^4x  \bigg[\sum_{j=0}^N e^{2 k j a} ( \partial^{\lambda} h_j^{\mu\nu})^2+\frac{1}{a^2} \sum_{j=0}^{N-1} e^{2 k j a} (h_{j}^{\mu\nu}-h_{j+1}^{\mu\nu})^2 \bigg] \nonumber \\
& = & -\frac 12 \int d^4x  \bigg[\sum_{j=0}^N ( \partial^{\lambda} h_j^{\mu\nu})^2+\frac{1}{a^2} \sum_{j=0}^{N-1} (h_{j}^{\mu\nu} - e^{-ka} h_{j+1}^{\mu\nu})^2 \bigg] ~,
\eea
where in the last term we rescaled the fields for canonical kinetic terms.  By comparing with \eq{pauli} we see that this is precisely the clockwork Pauli-Fierz mass term in transverse-traceless gauge, corresponding to
\beq
m_g =\frac{N}{\pi R} ~,~~~q_g = e^{\frac{k\pi R}{N}} ~.
\eeq
Thus clockwork gravity is realised as a deconstruction of gravity in the clockwork metric background.

\end{document}